\documentclass[12pt,preprint]{aastex}

\newcommand{\kms}{$\mbox{km s}^{-1}$}
\newcommand{\teff}{\mbox{$T_{eff}$}}
\newcommand{\hb}{\mbox{H$\beta$}}
\newcommand{\mgfe}{\mbox{[MgFe]$^\prime$}}
\newcommand{\naK}{\mbox{Na\,{\small I}}}
\newcommand{\caK}{\mbox{Ca\,{\small I}}}
\newcommand{\fea}{\mbox{Fe\,{\small I}~A}}
\newcommand{\feb}{\mbox{Fe\,{\small I}~B}}
\newcommand{\fe}{\mbox{$<$Fe\,{\small I}$>$}}
\newcommand{\mg}{\mbox{Mg\,{\small I}}}
\newcommand{\co}{\mbox{CO}}
\newcommand{\msun}{\mbox{M$_\sun$}}

\slugcomment{Submission version: 15 October 2007}

\shorttitle{K-band spectral indices}
\shortauthors{Silva, Kuntschner, Lyubenova}

\begin{document}

\title{A New Approach to the Study of Stellar Populations in
Early-Type Galaxies: K-band Spectral Indices and an Application to the
Fornax Cluster\altaffilmark{1}}

\altaffiltext{1}{Based on new observations performed at the European
Southern Observatory, Cerro Paranal, Chile; ESO program 68.B-0674A and
70.B-0669A as well as archival data from ESO program 076.B-0457A and
the SINFONI Science Verification dataset.}

\author{David R. Silva\altaffilmark{2}}
\affil{National Optical Astronomy Observatory\altaffilmark{3}}
\affil{950 North Cherry Ave., Tucson, AZ, 85748, USA}
\email{dsilva@tmt.org}

\altaffiltext{2}{Current assignment: Observatory Scientist, Thirty
Meter Telescope, 2632 East Washington Blvd., Pasadena, CA, 91107, USA}

\altaffiltext{2}{NOAO is operated by the Association of Universities
for Research in Astronomy (AURA), Inc., under cooperative agreement
with the National Science Foundation.}

\and

\author{Harald Kuntschner}
\affil{Space Telescope European Co-ordination Facility}
\affil{Karl-Schwarzschild-Str. 2, D-85748 Garching bei M\"unchen, Germany}
\email{hkuntsch@eso.org}

\and

\author{Mariya Lyubenova}
\affil{European Southern Observatory}
\affil{Karl-Schwarzschild-Str. 2, D-85748 Garching bei M\"unchen, Germany}
\email{mlyubeno@eso.org}

\begin{abstract}

The strong spectral features near 2.2 $\mu$m in early-type galaxies
remain relatively unexplored. Yet, they open a tightly focused window
on the coolest giant stars in these galaxies -- a window that can be
used to explore both age and metallicity effects. Here, new
measurements of K-band spectral features are presented for eleven
early-type galaxies in the nearby Fornax galaxy cluster. Based on
these measurements, the following conclusions have been reached:
(1) in galaxies with no signatures of a young stellar component, the
K-band \naK\ index is highly correlated with both the optical
metallicity indicator \mgfe\/ and the central velocity dispersion
$\sigma$;
(2) in the same galaxies, the K-band Fe features saturate in galaxies
with $\sigma > 150$ \kms\/ while \naK\/ (and \mgfe) continues to
increase;
(3) [Si/Fe] (and possibly [Na/Fe]) is larger in all observed Fornax
galaxies than in Galactic open clusters with near-solar metallicity;
(4) in various near-IR diagnostic diagrams, galaxies with signatures
of a young stellar component (strong \hb, weak \mgfe) are clearly
separated from galaxies with purely old stellar populations;
furthermore, this separation is consistent with the presence of an
increased number of M-giant stars (most likely to be thermally
pulsating AGB stars);
(5) the near-IR \naK\/ vs.~$\sigma$ or \fe\/ vs.~$\sigma$
diagrams discussed here seem as efficient for detecting putatively
young stellar components in early-type galaxies as the more commonly used
age/metallicity diagnostic plots using optical indices (e.g H$\beta$
vs.~\mgfe).
The combination of these spectral indices near 2.2 $\mu$m with high
{\it spatial} resolution spectroscopy from ground-based or space-based
observatories promises to provide new insights into the nature of
stellar populations in the central regions of distant early-type
galaxies.

\end{abstract}


\maketitle

\section{Introduction}
\label{sec:intro}

Understanding the stellar content of early-type galaxies is
fundamental to understanding their star formation and chemical
evolution history.  Most early-type galaxies are too distant to
resolve their individual stars with current technology, rendering the
direct study of their stellar populations impossible.  Thus, their
stellar populations must be studied using indirect methods.

In recent decades, significant effort has gone into trying to better
constrain the stellar contents for early-type galaxies using optical
spectroscopic data. The most commonly studied features have been Ca I
H and K 0.38 $\mu$m, H$\beta$, Mgb 0.52 $\mu$m, Fe $\mu$m 0.53, Na
0.82 $\mu$m, and CaT 0.86 $\mu$m. Interpretation of all such spectral
features is intrinsically complicated by their blended nature -- each
feature is really the super-position of many spectral lines, usually
from several different elements, blurred together by the line-of-sight
velocity dispersion within each galaxy. There is no way to overcome
this problem -- it must simply be taken into account during analysis.
As population synthesis models have become more sophisticated and
digital stellar libraries more complete, this problem has become more
tractable over time.

Another challenge arises from the composite nature of galaxies: each
observed feature is the luminosity-weighted integrated sum of that
feature from all stars in the observed line-of-sight.  Naturally,
luminosity-weighted does not imply mass-weighted. A relatively small
fraction of the mass can dominate the observed luminosity and mask the
underlying stellar population (e.g. as happens during a starburst
event within a pre-existing galaxy).

Even in relatively quiescent galaxies, light from stars at several
important evolutionary stages contribute roughly equally to the
observed spectral features between 0.4 -- 1 $\mu$m range. Hence, a
feature depth change could be due to (e.g.) a change near the (mostly)
age-driven main-sequence turnoff or the (mostly) metallicity-driven
red giant branch.  The details can become quite complicated, as
illustrated by the long standing controversy about whether observed
changes in Balmer line strength arise from the presence of younger
main sequence stars, more metal-poor main sequence stars, or an
extended horizontal giant branch (for recent discussions of this
debate, see Maraston \& Thomas 2000 and Trager et al. 2005). A similar
controversy surrounds Na 0.82 $\mu$m feature: is it driven by
metallicity-driven red giant branch changes, initial mass function
related differences in the relative number of cool dwarf and giant
stars or both \citep[e.g.][]{car86, all89, del92}?

However, the properties of the RGB component can be isolated by
observing in the K-band (centered near 2.2 $\mu$m).  At those
wavelengths, cool giants near the tip of the first-ascent red giant
branch (RGB) dominate the integrated light in old ($\geq$ 3 Gyr)
stellar populations. In combination with optical observations, K-band
observations should facilitate the separation of MSTO and RGB light
contributions. There are two possible complications to this scenario.
First, a very young stellar population containing red supergiants will
contribute a significant fraction K-band light. Fortunately, such a
population is obvious from the presence of H~II region emission lines
at shorter wavelengths. Second, a somewhat older population (1 -- 2
Gyr, i.e. an intermediate-age population) may contain bolometrically
bright carbon stars that can contribute a detectable amount of K-band
light (see discussions in Silva \& Bothun 1998a,b). Such a population
may or may not be connected to increased H$\beta$ strength.

Initial development of these ideas can be found in Silva et
al. (1994), Mobasher \& James (1996), James \& Mobasher (1999),
Mobasher \& James (2000), all of whom focused on the CO 2.36 $\mu$m
feature. Origlia et al. (1997) observed a Si dominated feature at
1.59 $\mu$m as well as CO dominated features. These observational
studies were limited by small-format detectors to relatively low
resolving powers and/or small wavelength ranges per observation. In
the cases of Silva et al. and Origlia et al., only small,
heterogeneous samples of galaxies were observed. A general conclusion
of the James \& Mobasher studies was that changes in CO strength
between early-type galaxies in high-density and low-density regions
were statistically consistent with different fraction contributions of
intermediate-age AGB light and hence galaxies in low-density regions
had younger luminosity-weighted mean ages. Origilia et al. argued that
[Si/Fe] was super-solar in the four elliptical galaxies they observed.

To further develop these ideas and investigate the usefulness of other
K-band spectral indices in the study of early-type galaxies, new data
have been obtained for eleven E/S0 galaxies in the nearby Fornax
cluster. Only measurements in the central regions of these galaxies
are discussed here.  In Section~\ref{sec:data}, the galaxy sample and
its observations are discussed, while in Section~\ref{sec:proc} the
data processing methodology is described. The measurement of spectral
feature strength is explained in Section~\ref{sec:lines} while basic
observation results are presented in Section~\ref{sec:results}. The
broader astrophysical implications of our observational results are
discussed in Section~\ref{sec:disc}. A summary is provided at the end.

\section{Observations}
\label{sec:data}

Long-slit spectroscopic data obtained with ISAAC at the ESO Very Large
Telescope (VLT) have been combined with data obtained with SINFONI at
the VLT to study a small sample of early-type galaxies in the nearby
Fornax cluster. Details about the sample as well as the instrumental
setups for these two instruments are presented in this section.

\subsection{Samples}

The full sample observed consists of eleven early-type galaxies in the
nearby Fornax cluster, sub-divided into six elliptical galaxies and
five S0 galaxies. The overall sample was selected to cover a
significant range in luminosity weighted age, metallicity and
mass. Basic galaxy properties are summarized in
Table~\ref{tab:galSample}.

\begin{figure}
\resizebox{\hsize}{!}{\includegraphics{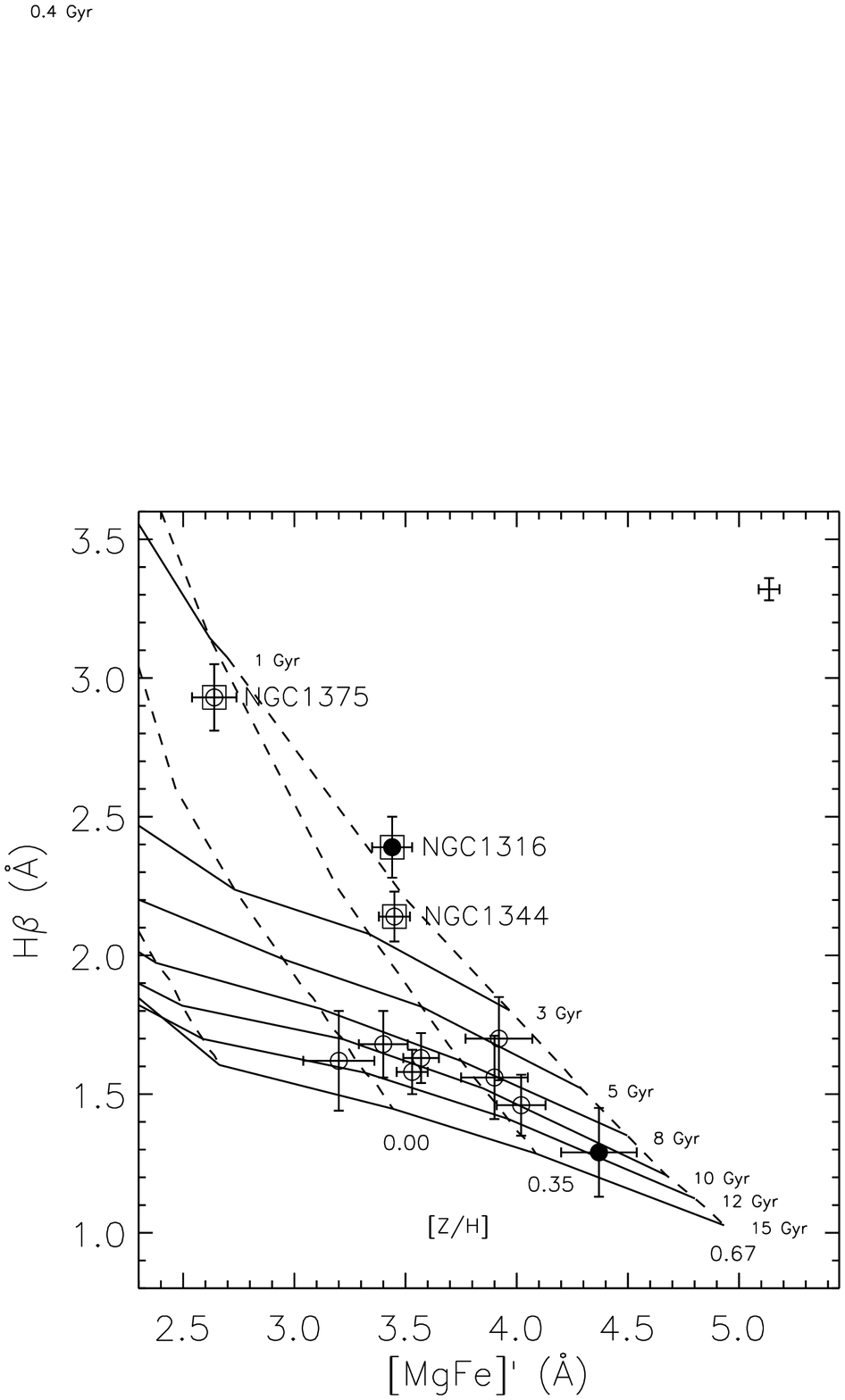}}
\caption{\label{fig:fornax_pops} \hb\/ versus \mgfe\/ age/metallicity
diagnostic diagram. Overplotted are simple stellar population models
for solar abundance ratios by Thomas, Maraston \& Bender (2003). Solid
and dashed lines show lines of constant luminosity weighted age and
metallicity, respectively. Filled circles represent observations for
NGC\,1316 and NGC\,1399 for which the IR indices were obtained with
VLT-SINFONI and are extracted from a 2$\times$3\arcsec\/ aperture. For
all other galaxies (open circles) 1/8 effective radius extractions are
shown. Galaxies with a significant contribution of an intermediate age
(1--2\,Gyr) population are marked by open squares. The error bar in
the upper right corner indicates the uncertainty in the transformation
to the Lick/IDS system. For details see
Section~\ref{sec:proc-optical}.  }
\end{figure}

Optical long-slit spectroscopy is available for all observed galaxies
(Kuntschner 2000, Kuntschner et al. 2002) allowing comparison between
the K-band line-strength indices derived in this paper with optical
indices. When compared to simple stellar population models, three
galaxies (NGC\,1316, NGC\,1344 and NGC\,1375) appear to have a
significant contribution from young stars resulting in luminosity
weighted ages between 1 and 3\,Gyrs (see
Figure~\ref{fig:fornax_pops}). The remaining galaxies are consistent
with being dominated by old stellar populations with a range in
average metallicity.

Multi-color HST Advanced Camera for Survey images (courtesy of
A. Jorden) of all the galaxies in our sample were inspected for the
presence of dust. Three galaxies (NGC 1316, NGC 1344, and NGC 1380 --
see details below) have clear central dust features. The other
galaxies in our sample have no visible dust feature within the
investigated aperture.

%
%

A few additional object specific remarks seem warranted:

\begin{itemize}

\item[]{\it NGC 1316} -- Fornax A, S0(3)pec. This galaxy has large
scale, irregular, central dust features (based on inspection of
HST-ACS images).

\item[]{\it NGC 1344} -- this S0 galaxy contains morphological shells
in its outer regions \citep{mal80} as well as optical spectral
signatures of a younger central population (e.g. Kuntschner
et al. 2002). Inspection of HST-ACS images reveals patchy dust within the
central 5 arcsec. In combination, these characteristics suggest (1 --
2 Gyr) dynamical interaction or merger event that triggered a central
star formation episode.\smallskip

\item[]{\it NGC 1375} -- Kuntschner (2000) concluded that the mean age
of the central population of this S0 galaxy is significantly lower
than for the other galaxies in our sample.\smallskip

\item[]{\it NGC 1380} -- contains a kinematically decoupled core (KDC)
\citep{lon94}. It also contains an optically thick,
well-organized central dust lane, easily visible in available HST
images.\smallskip

\item[]{\it NGC 1381} -- this S0 galaxy contains X-bulge isophotal
pattern, often attributed to a recent dynamical encounter or merger
event.\smallskip

\item[]{\it NGC 1399} -- cD, E0

\item[]{\it NGC 1419} -- this E0 galaxy has anomalously bright K-band
surface brightness fluctuations (SBF) relative to its I-band surface
brightness fluctuations, suggesting the presence of extended giant
branches, perhaps due to the presence of an intermediate-age
population \citep{mei01, liu02}.\smallskip

\item[]{\it NGC 1427} -- this E4 galaxy also has anomalously bright
K-band SBF (see NGC 1419 SBF references above). It also contains a KDC
\citep{don95}.

\end{itemize}

%
%

\begin{deluxetable}{llcccccccl}
\rotate
\tablewidth{0pt}
\tablecolumns{10}
\tabletypesize{\small}
\tablecaption{\label{tab:galSample} Galaxy Sample}

\tablehead{
\colhead{Galaxy}
&\colhead{Type}
&\colhead{\it z}
&\colhead{$\sigma$}
&\colhead{R$_e$ maj}
&\colhead{R$_e$ min}
&\colhead{PA maj}
&\colhead{R$_e$ extr.}
&\colhead{SNR}
&\colhead{Note}
\\
 \colhead{}
&\colhead{}
&\colhead{}
&\colhead{(\kms)}
&\colhead{(\arcsec)}
&\colhead{(\arcsec)}
&\colhead{(\arcdeg)}
&\colhead{(\arcsec)}
&\colhead{}
&\colhead{}
\\
 \colhead{(1)}
&\colhead{(2)}
&\colhead{(3)}
&\colhead{(4)}
&\colhead{(5)}
&\colhead{(6)}
&\colhead{(7)}
&\colhead{(8)}
&\colhead{(9)}
&\colhead{(10)}
}

\startdata
NGC 1316 & So(3)pec    & 0.0059 & 226 & 132.2  &  90.5 &   50 &  \nodata & 168 & Fornax A; SINFONI data \\
NGC 1344 & E5          & 0.0039 & 171 &  26.7  &  15.6 &  165 &  3.3     & 121 & Type from RC3 \\
NGC 1374 & E0          & 0.0043 & 196 &  26.6  &  24.2 &  118 &  3.3     & 124 & KDC \\
NGC 1375 & S0(cross)   & 0.0025 &  70 &  24.0  &   9.3 &   91 &  3.0     & 48  & Central starburst \\
NGC 1379 & E0          & 0.0044 & 138 &  23.9  &  23.3 &  182 &  2.9     & 78  &\nodata \\
NGC 1380 & S0/a        & 0.0063 & 213 &  52.4  &  26.6 &    7 &  3.4     & 213 &\nodata \\
NGC 1381 & S0(9)(boxy) & 0.0057 & 169 &  23.5  &   8.0 &  139 &  2.9     & 135 &\nodata \\
NGC 1399 & E0          & 0.0048 & 360 & 134.1  & 120.8 &  110 &  \nodata & 111 & cD; SINFONI data \\
NGC 1404 & E2          & 0.0065 & 221 &  26.0  &  23.0 &  157 &  157     & 280 & \nodata \\
NGC 1419 & E0          & 0.0071 & 128 &   9.4  &   9.3 &  188 &  1.2     & 111 & K-band SBF outlier \\
NGC 1427 & E4          & 0.0046 & 186 &  39.2  &  27.1 &   76 &  4.9     & 104 & K-band SBF outlier, KDC \\
\enddata

\tablecomments{ Notes: Galaxy types (column 2) were extracted from
\citet{fer89} while redshifts (column 3) are taken from the NASA
Extragalactic Database (NED) -- original redshift references:
\citep{dac98}, \citet{don95}, \citet{gra98} and \citet{dev91}. Central
velocity dispersions (column 4) are taken from our own measurements of
the ISAAC and SINFONI data (see Sections~\ref{sec:isaac_gals} and
\ref{sec:proc-sinfoni}). The values for the effective radii (columns 5
and 6) were taken from \citet{cao94} and derived from the RC3 for
NGC\,1344.  The position angle of the major axis (column 7) was taken
from the RC3 and \citet{car91}. The 1/8 R$_e$ extraction radius which
was used do extract a central spectrum for the ISAAC observations is
given in column 8. Empirical SNR estimates for the extracted spectra
are given in column 9 (see Section 3.1 for details). Comments are
given in column 10.}

\end{deluxetable}

A small sample of Galactic open cluster stars were also observed.
Their basic properties are summarized in Table~\ref{tab:starSample}.
Cluster parameters (age, metallicity) and star photometry come from
Houdashelt, Frogel, \& Cohen (1992). Age has units of Gyr. Spectral
types (SpT) have been assigned based on their $J-K$ colors following
\citet{joh66} for K giant stars and \citet{lee70} for M giant
stars. These stars have several roles: velocity dispersion templates,
CO 2.36 $\mu$m index comparisons, and small independent framework for
interpreting K-band line strength variations as functions of basic
atmospheric parameters.  Each of these roles will be discussed in turn
below.

%
%

\begin{deluxetable}{llllccl}
\tablewidth{0pt}
\tablecolumns{7}
\tabletypesize{\small}
\tablecaption{\label{tab:starSample} Galactic open cluster stars}

\tablehead{
\colhead{Cluster}
&\colhead{Age}
&\colhead{[Fe/H]}
&\colhead{Star}
&\colhead{$V-K$}
&\colhead{$J-K$}
&\colhead{SpT}
\\
\colhead{(1)}
&\colhead{(2)}
&\colhead{(3)}
&\colhead{(4)}
&\colhead{(5)}
&\colhead{(6)}
&\colhead{(7)}
}

\startdata
NGC 2204 & 2.8 & --0.38 
           & 1146 & 5.71 & 1.12 & M4 \\
         &&& 3304 & 3.10 & 0.79 & K3 \\
         &&& 3325 & 3.84 & 0.94 & M0.5 \\
         &&& 4132 & 5.03 & 1.11 & M3.5 \\
\hline
NGC 2477 & 1.0 & --0.02 
           & 1069 & 4.07 & 0.95 & M1 \\
         &&& $\lambda$ & 2.96 & 0.71 & K3 \\
         &&& 2117 & 4.18 & 0.96 & M2.5 \\
         &&& 6053 & 2.81 & 0.68 & K2 \\
\hline
NGC 2506 & 3.5 & --0.52  
           & 2401 & 3.65 & 0.91 & K5 \\
         &&& 4228 & 4.07 & 0.99 & M1 \\
\enddata
\end{deluxetable}

Finally, several bright hot stars were observed to measure and correct
the telluric absorption features in the K-band.  B stars were selected
by color from the Hipparcos Tycho catalog (ESA, 1997). In the spectral
range of interest, these stars have no spectral features; therefore,
that any observed features must be telluric absorption lines. In
addition to color, proximity to the target objects and brightness were
used as secondary selection criteria.  The actual stars observed are
listed in Table~\ref{tab:telluric}. The {\it V} magnitude and {\it
B--V} color are from the Tycho catalog (ESA, 1997).

%
%

\begin{deluxetable}{llcc}
\tablewidth{0pt}
\tablecolumns{4}
\tabletypesize{\small}
\tablecaption{\label{tab:telluric} ISAAC telluric correction stars}

\tablehead{
\colhead{Telluric Star} &
\colhead{Target} &
\colhead{$V$} &
\colhead{$B-V$}
\\
\colhead{(1)} &
\colhead{(2)} &
\colhead{(3)} &
\colhead{(4)} 
}

\startdata
Tycho2 7034 01323 & Fornax galaxies & 6.07 & --0.11 \\
Tycho2 5942 02406 & NGC 2204 & 5.29 & --0.19 \\
Tycho2 7645 03250 & NGC 2477 & 6.13 & --0.17 \\
Tycho2 5425 00772 & NGC 2506 & 7.87 & --0.14 \\
\enddata
\end{deluxetable}

\subsection{ISAAC observations}
\label{sec:obs-isaac}

Most of the observations discussed in this paper were obtained at the
ESO Very Large Telescope using the ISAAC near-IR imaging spectrometer
mounted at one of the Nasmyth foci of the 8.2m UT1/Antu
telescope. Galaxy observations were obtained in 2002 January and 2002
November.  A summary observations log is given in
Table~\ref{tab:obsLog}. For each galaxy, the following information is
provided: UT Date of observation, integration time per frame, number
of frames, and total integration time. Since NGC 1380 and NGC 1404
were observed multiple times, each observation sequence is assigned a
sequence identifier (A, B, C, ...).

%
%
\begin{deluxetable}{llccccl}
\tablewidth{0pt}
\tablecolumns{4}
\tabletypesize{\small}
\tablecaption{\label{tab:obsLog} ISAAC Galaxy Observations Log}

\tablehead{
\colhead{UT Date} &
\colhead{Galaxy} &
\colhead{$\tau$} &
\colhead{Frames} &
\colhead{Total $\tau$} &
\colhead{PA} &
\colhead{SeqID} 
\\
\colhead{} &
\colhead{} &
\colhead{(secs)} &
\colhead{} &
\colhead{(secs)} &
\colhead{} &
\colhead{} 
\\
\colhead{(1)} &
\colhead{(2)} &
\colhead{(3)} &
\colhead{(4)} &
\colhead{(5)} &
\colhead{(6)} &
\colhead{(7)} 
}

\startdata
2002 Jan 02 & NGC 1404 & 500 & 6 & 3000 & 89.45 & A \\
2002 Jan 14 & NGC 1404 & 500 & 6 & 3000 & \nodata & B \\
2002 Jan 14 & NGC 1404 & 500 & 6 & 3000 & \nodata & C \\
2002 Nov 12 & NGC 1404 & 500 & 6 & 3000 & \nodata & D \\
2002 Nov 13 & NGC 1404 & 500 & 6 & 3000 & \nodata & E \\
&&&&& \\
2002 Nov 13 & NGC 1380 & 400 & 8 & 3200 & 89.45 & A \\
2002 Nov 13 & NGC 1380 & 400 & 8 & 3200 & \nodata & B \\
2002 Nov 12 & NGC 1380 & 400 & 8 & 3200 & \nodata & C \\
&&&&& \\
2002 Nov 12 & NGC 1344 & 400 & 8 & 3200 & 164.45 & \nodata\\
2002 Nov 13 & NGC 1374 & 400 & 8 & 3200 & 89.45 & \nodata\\
2002 Nov 12 & NGC 1375 & 400 & 8 & 3200 & 89.45 & \nodata\\
2002 Nov 13 & NGC 1379 & 400 & 8 & 3200 & 89.45 & \nodata\\
2002 Nov 13 & NGC 1381 & 400 & 8 & 3200 & 138.45 & \nodata\\
2002 Nov 12 & NGC 1419 & 400 & 8 & 3200 & 89.45 & \nodata\\
2002 Nov 12 & NGC 1427 & 400 & 8 & 3200 & 77.45 & \nodata\\
\enddata
\end{deluxetable}

%
%
\begin{deluxetable}{cccccc}
\tablewidth{0pt}
\tablecolumns{6}
\tabletypesize{\small}
\tablecaption{\label{tab:obsSinfoni} SINFONI Galaxy Observations Log}
\tablehead
{
\colhead{UT Date} &
\colhead{Galaxy} &
\colhead{$\tau$} &
\colhead{O$+$S Frames} &
\colhead{Total $\tau$} &
\colhead{SeqID}
\\
\colhead{} &
\colhead{} &
\colhead{(secs)} &
\colhead{} &
\colhead{(secs)} &
\colhead{} 
\\
\colhead{(1)} &
\colhead{(2)} &
\colhead{(3)} &
\colhead{(4)} &
\colhead{(5)} &
\colhead{(6)} 
}
\startdata
2004 Nov 25 & NGC\,1399 &  300 & 4$+$4 & 1200 & B \\
2004 Nov 26 & NGC\,1399 &  300 & 4$+$4 & 1200 & C \\
2004 Nov 26 & NGC\,1399 &  300 & 4$+$4 & 1200 & D \\
2004 Nov 26 & NGC\,1399 &  300 & 4$+$4 & 1200 & E \\
&&&&& \\
2005 Oct 13 & NGC\,1316 &  600 & 3$+$1 & 1800 & A \\
2005 Oct 13 & NGC\,1316 &  600 & 4$+$2 & 2400 & B \\
\enddata
\end{deluxetable}

ISAAC was used in short-wavelength spectroscopic mode at
medium-resolution (SWS-MR). In this mode, the array used is a
1024$\times$1024 Hawaii Rockwell HgCdTe array with a spatial plate
scale of 0\farcs148 pix $^{-1}$. A 120$\times$1\arcsec\/ slit was
used. The dispersion and spectral resolution measured from our arc
lamp data were 1.21\,\AA\ pix$^{-1}$ and 7.7\,\AA\/ (FWHM, $R \simeq
2900$), respectively.  The approximate wavelength range was 2.12 --
2.37 $\mu$m but the central wavelength was always corrected for object
redshift to ensure that all relevant spectral features were observed.

Galaxy observations were obtained using the standard nod-on-slit mode.
At the start of each observational sequence, the galaxy was centered on
the slit near one end and a individual integration was executed.  The
galaxy was then moved approximately 60\arcsec\/ towards the other end
of the slit where two more integrations were executed. The galaxy was
then returned to the original slit position where another integration
was obtained. This ABBA pattern was repeated a number of times,
resulting in multiple individual two-dimensional spectroscopic images.
%
%
Including time for telescope offsetting and array readout, each
sequence took roughly one hour to execute.

After each set of galaxy sequences was completed, telluric standard
Tycho2 7034 01323 was observed (see Table~\ref{tab:telluric}) at five
different positions along the slit.  At each position, a short ($\leq$
10 secs) integration was obtained. At the end of the telluric
integrations, a XeAr lamp-on/lamp-off sequence was obtained followed by
a quartz lamp-on/lamp-off sequence. The spectrograph setup was not
changed during the entire galaxy-standard-lamps sequence, ensuring that
the grating did not move. Thus, each star-lamp sequence had the same
central wavelength as the corresponding galaxy sequence. This is
critical for proper correction of telluric features, dispersion, and
array pixel-to-pixel response.

On various dates, stars from the following open clusters were
observed: NGC 2204 (2002 November 12), NGC 2477 (2001 December 23),
and NGC 2506 (2002 November 13). The actual stars observed are listed
in Table~\ref{tab:starSample}. Each star was observed at five (5)
positions along the slit.  At each position, a short ($\leq$ 10 secs)
integration was obtained. After each set of cluster star observations,
a telluric standard was observed, followed by arc and flat lamp
observations (as described above). The exact telluric standards
observed are listed in Table~\ref{tab:telluric}.

As part of the VLT calibration program, calibration data was also
obtained during the day following each set of observations. These data
include observations of quartz and arc lamps (for flat-field and
wavelength dispersion corrections) as well as dark frames with the
same detector integration times (DITs) as the night-time science
observations. For this project, the daytime quartz and arc lamps were
not useful due to grating angle positioning inaccuracy.

\subsection{SINFONI observations}
\label{sec:obs-sinfoni}

Observations of NGC\,1316 (Fornax A) and NGC\,1399 (the central cD
galaxy) were obtained with SINFONI, a near-IR integral field unit (IFU)
spectrograph mounted at the Cassegrain focus of VLT UT4 (Yepun). A
spatial scale of the 0.1$\arcsec$ scale was used, corresponding to a
total field of view of approximately $3\arcsec \times 3\arcsec$. The
spectra cover the wavelengths between 1.95 -- 2.4 $\mu$m at
2.45\,\AA/pix. The spectral resolution is 6.1\,\AA\/ FWHM ($R \simeq
3700$).

NGC\,1399 was observed 2004 November 25 and 29 during the SINFONI
Science Verification program (Eisenhauer et al. 2003; Bonnet et
al. 2004). The NGC\,1316 data come from the ESO Science Archive
(Program 076.B-0457A, PI Ralf Bender). The original observations were
made on 2005 October 13. Details of these observations are given in
Table~\ref{tab:obsSinfoni}.

NGC\,1399 was observed in a sequence OSSOOSSO (O = object integration,
S = sky/background integration) to enable accurate background
correction. NGC\,1316 was observed with a slightly different pattern
-- OSOOSO. As with the ISAAC observations discussed above, telluric
and velocity template standard stars were observed periodically. The
former were hot stars while the latter covered the range K5III to
M1III.

\section{Data Processing}
\label{sec:proc}

\subsection{ISAAC data reduction}
\label{sec:proc-isaac}

Data processing was done using a combination of shell scripts, {\tt
eclipse} \citep{dev97}, and IRAF\footnote{The Image Reduction and
Analysis Facility (IRAF) is distributed by the National Optical
Astronomy Observatories (NOAO), which are operated by the Association
of Universities for Research in Astronomy, Inc. (AURA) under contract
with the National Science Foundation (NSF).}.

\subsubsection{Quartz and Arc Lamps}

As described above, each ISAAC observing sequence consists of a set of
galaxy or cluster star observations followed immediately by associated
telluric star and lamp exposures. The lamps exposures were processed
first. The resultant flat-field and XeAr arc lamp frames were used in
turn to process their associated telluric star and galaxy or cluster
star sequences.

A raw flat-field frame was constructed by subtracting the lamp-off
exposure from the lamp-on exposure to remove thermal background and
dark current.  The difference frame was used to define the illuminated
region of the ISAAC SW detector (the entire detector is not
illuminated in SWS mode). The difference frame was trimmed to remove
the unilluminated part. The trimmed frame was then collapsed in the
spatial dimension into a one-dimensional vector. A one-dimensional
polynomial was fit to this vector. This polynomial was then divided
into every row of the original background-corrected flat-field frame
frame. The resultant frame is suitable for removing pixel-to-pixel
sensitivity variations.  These flat-field frames obtained immediately
after all telluric sequences are preferred to evening or morning flats
since tests demonstrated that such so-called daytime calibrations did
not accurately remove fringing. This is presumably due to grating
re-positioning imprecision.

The XeAr exposures were processed by first subtracting the arc-off
frame from the arc-on frame, again to remove any background
illumination not associated with the XeAr lamps. A two-dimensional
wavelength dispersion correction was derived, using low-order
polynomials in both dimensions.

\subsubsection{Telluric Correction Star Observations}

For each telluric absorption correction star sequence, the five
individual frames were first trimmed to correspond to the illuminated
part of the detector (see above) and then divided by their associated
flat-field.  A background$+$dark frame was determined by median
combining the individual flattened exposures.  This background frame
was then subtracted from those exposures.  The five flat-field and
background corrected frames were summed into a single frame.  A
spectral tilt correction was determined from the resultant
frame. Spectral tilt and wavelength dispersion corrections were then
applied to the summed, flat-field corrected frame.  The five
individual spectra were extracted and averaged into a single spectrum
using a sigma-clipping algorithm for bad pixel rejection. This
combined spectrum was used to produce two things: a normalized
telluric absorption line correction spectrum and a instrument
sensitivity correction.

Building the normalized telluric correction spectrum started by
fitting a one-dimensional bi-cubic spline to the telluric star
spectrum. Care was taken to reject high outliers (e.g. cosmic rays,
bad pixels) but not too many low outliers (likely to be the telluric
absorption features of interest). The derived vector was divided back
into the original spectrum to normalize it and remove the continuum
shape.

This bi-cubic spline fit was then re-used to determine an instrument
sensitivity curve.  Since the color of the telluric stars is known
from the Hipparcos catalog, a reasonable estimate of their effective
temperature can be made.  Furthermore, since these are hot stars, we
know that their continuum in the K-band is well approximated by the
Rayleigh-Jeans part of the blackbody spectrum associated with their
effective temperature. By computing the appropriate blackbody curve in
the relevant wavelength range and then dividing it by the bi-cubic
spline fit (i.e. a representation of the observed stellar continuum),
the instrument sensitivity function can be estimated. Of course, this
is not an absolute correction -- but an absolute correction is not
necessary for the spectral line strength measurements described below.

\subsubsection{Cluster Star Observations}

Cluster star sequences were processed into single combined spectra in
a similar manner as telluric star sequences.  At that point, the
combined spectrum of each star was divided by its associated
normalized telluric absorption line correction spectrum. To remove the
instrument response signature, the telluric corrected cluster star
spectrum is then multiplied by the instrument sensitivity function.

To achieve the optimal telluric absorption line correction, it was
typically necessary both to shift the telluric star spectrum in the
dispersion direction (usually by less than 1 pixel) and scale it
(usually by a few percent) before the division. The physical cause of
the physical shift in the dispersion direction is unclear. However,
the need to scale is simply due to the slight changes in airmass and
time between object and telluric correction standard observations. It
is also clear that the strength of every telluric absorption feature
in this wavelength window is not changing in concert. Thus, the
telluric correction is not perfect at all wavelengths. In principle, a
more sophisticated correction process is possible. In practice, the
current level of correction is good enough for this application.

\begin{figure}
\resizebox{\hsize}{!}{\includegraphics[angle=0]{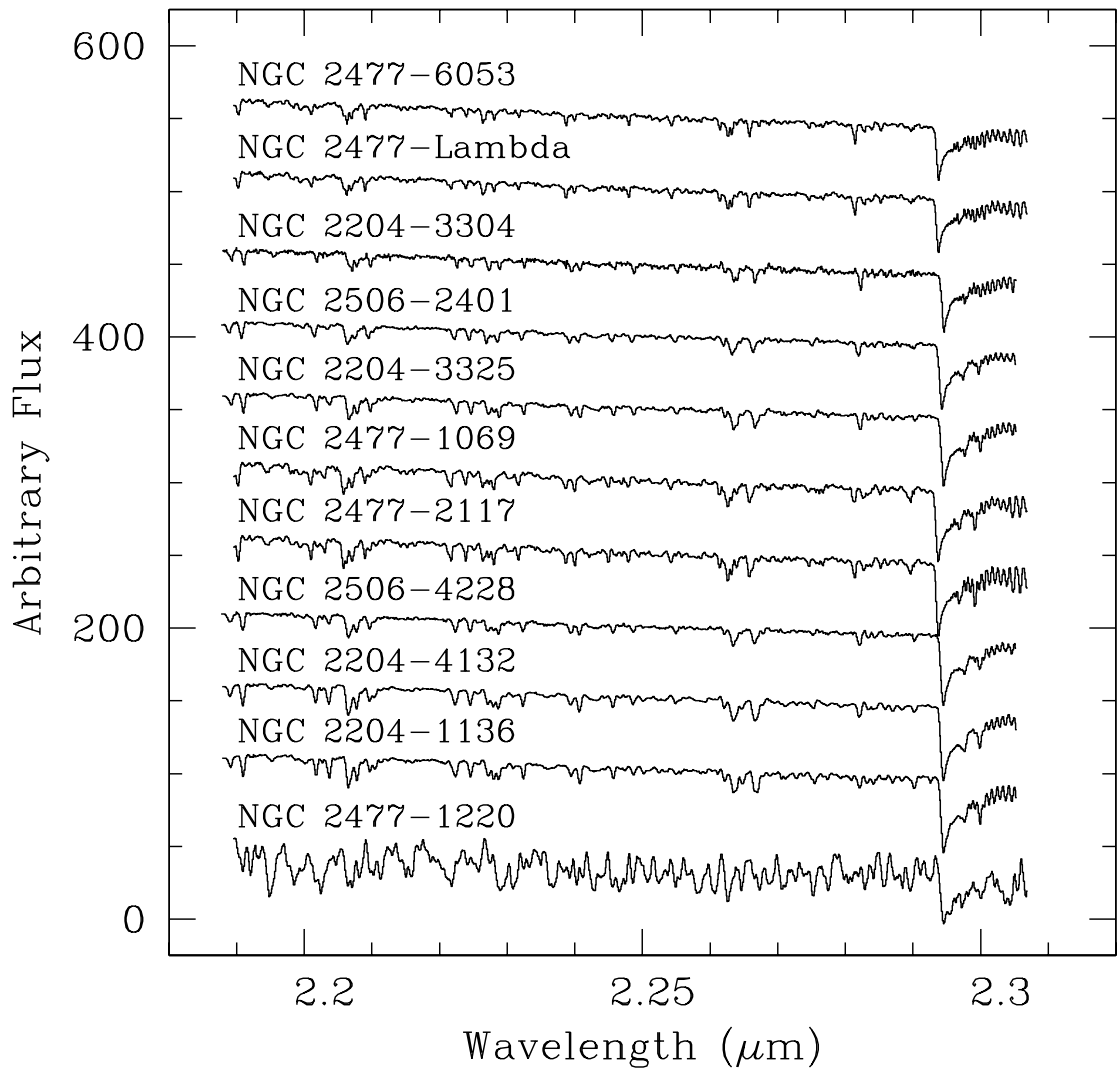}}
\caption{\label{fig:clusterStars} Galactic Open Cluster Stellar
  Spectra -- Final spectra for all the stars listed in
  Table~\ref{tab:starSample}. Each spectrum was normalized to 100 at
  2.25 $\mu$m and then offset by a fixed amount for illustration
  purposes. The spectra are ordered (top to bottom) by increasing
  $J-K$ color (hence, decreasing effective temperature). Note how the
  \caK\/ 2.207 $\mu$m and \naK\/ 2.265 $\mu$m atomic features as well
  as the \co\/ 2.36 $\mu$m molecular feature increase in depth with
  decreasing effective temperature. Star NGC\,2477-1220 is a carbon
  star and not discussed further in this paper.}
\end{figure}

The final processed spectra for all observed cluster stars are shown
in Figure~\ref{fig:clusterStars}.

%
%

\subsubsection{Galaxy Observations}
\label{sec:isaac_gals}

The main technical goal of this paper is to make accurate line strength
measurements in the central regions of our target galaxies. However, in
the future, the radial line strength gradients will also be studied.
Since each AB galaxy observation is only separated by $\sim$
100\arcsec, a simple AB subtraction (i.e. the typical procedure in the
near-IR) could cause inaccurate dark and background correction in the
lower surface brightness parts of the galaxy luminosity profiles.
Therefore, a procedure derived from the processing of {\it optical}
long-slit spectra was applied.

\begin{figure}
\resizebox{\hsize}{!}{\includegraphics[angle=0]{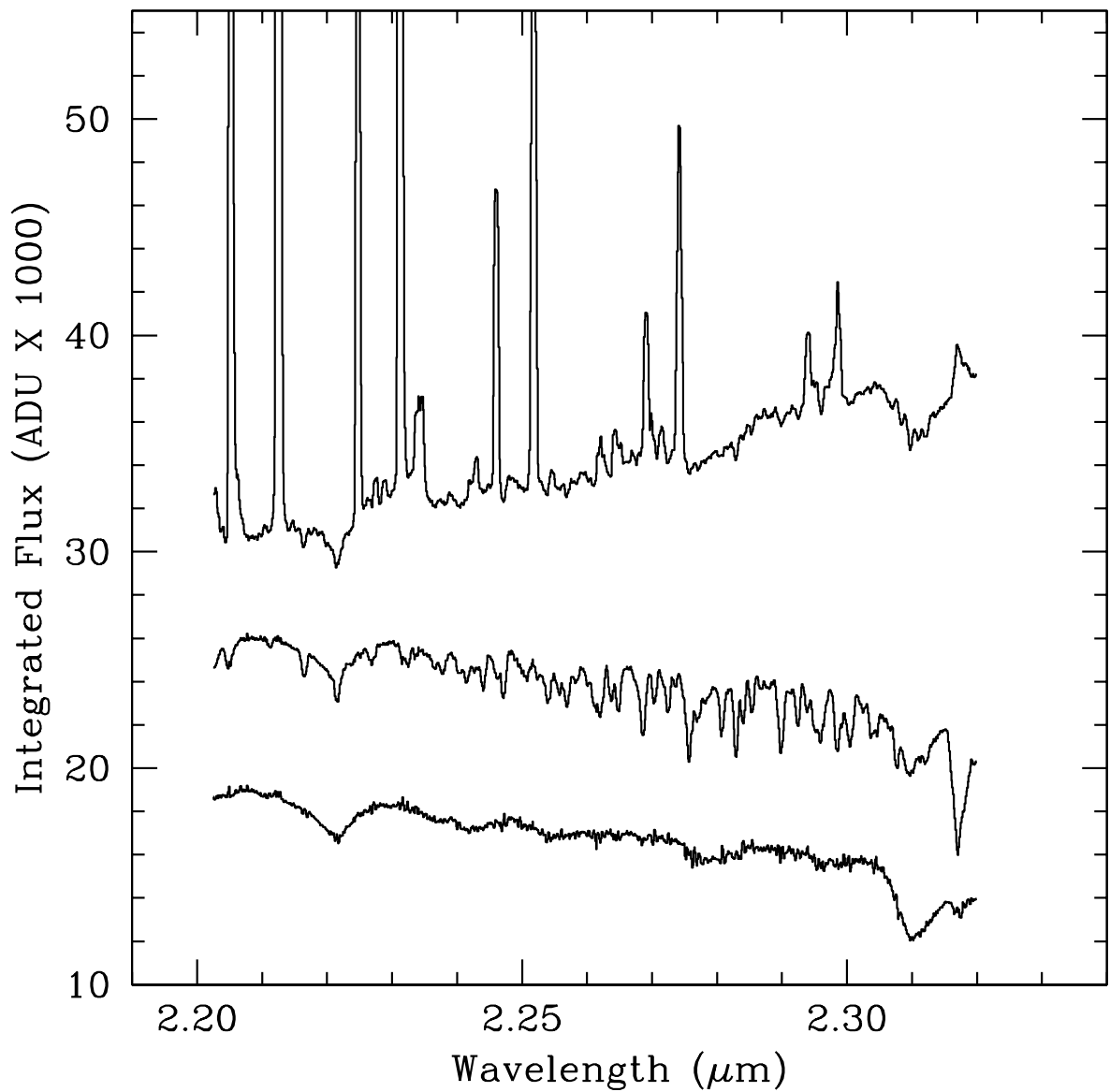}}
\caption{\label{fig:galProcExample} Galaxy Processing Example -- A
  central 3\arcsec\/ diameter extraction of NGC 1404 (Sequence D, see
  Table~\ref{tab:galSample}) are shown (from top-to-bottom): before
  corrections for dark current, background, telluric absorption, and
  instrument sensitivity; after dark current and background (emission)
  correction; and after correction for telluric absorption and
  instrument sensitivity.  No correction for redshift has been applied.
  The top and bottom spectra are shown at their observed relative
  integrated flux levels. Some of the telluric emission lines have
  integrated fluxes greater than 80 000 ADU. The middle and bottom
  spectra have been separated by an arbitrary additive offset;
  otherwise, they would lie on top of each other. }
\end{figure}

The basic data processing challenge is shown in
Figure~\ref{fig:galProcExample} where a central 3\arcsec\/ diameter
extraction of NGC\,1404 (sequence D; see Table~\ref{tab:galSample}) is
shown at three stages: before background correction, after background
correction but before telluric absorption line correction, and after
all processing.  One of the major advantages of ISAAC over other
near-IR spectrometers is the ability to use a high-enough spectral
resolution to resolve telluric features so that they can be accurately
removed.

%
%

As shown in Table~\ref{tab:galSample}, each galaxy observation
consists of a series of 6 or 8 individual exposures taken in ABBA
sequences. To begin, each galaxy frame was trimmed and the appropriate
median dark frame was subtracted. Following the advice of the {\it
ISAAC Data Reduction Guide} (available on-line from the ESO ISAAC Web
page), electronic ghosting was removed using the {\tt eclipse} recipe
{\it isaacp ghost}.

Recall that each galaxy sequence had an associated telluric$+$lamp
sequence.  As described in the telluric star processing section above,
each telluric$+$lamp sequence was processed into a series of
calibrations: flat-field, two-dimensional wavelength dispersion
solution, spectral tilt correction, normalized telluric absorption
correction vector, and instrument sensitivity function. After
electronic ghost correction, each individual galaxy frame was divided
by this flat-field and then geometrically rectified using the derived
dispersion and spectral tilt solutions. In short, the galaxy sequence
and its associated telluric star are calibrated using the same frames
and vectors.

A slit illumination correction was applied next. Since twilight sky
exposures were not available, the correction function had to be
reconstructed from the actual observations. In each individual galaxy
observation, the spatial dimension can be divided into a galaxy
sub-section and a sky sub-section. For each AB pair, the sky
sub-section is on opposite sides of the slit center. The A and B sky
sub-sections were extracted and merged into a single A$+$B sky frame,
after applying a small additive offset to the B sky sub-section to
align it better with the A sky sub-section. This offset is caused by
temporal changes in the background between the A and B exposures.
This process was repeated for all A and B galaxy exposures on a given
night.  The resultant frames were then scaled to a common mean value
and averaged using a sigma clipping bad pixel rejection scheme.  That
average sky frame was collapsed in the dispersion direction into a
one-dimensional vector. A nightly illumination correction function was
derived by fitting a second-order bi-cubic spline to that vector and
normalizing the result. This illumination correction was then divided
into all galaxy frames from that night.

The next step was to remove the background. The standard procedure
would be to select a radial position dominated by the background,
determine the mean background at this position, and subtract that
value at every pixel in the spatial dimension. As applying this
procedure to the current dataset did not produce satisfactory results,
an alternative two-step approach was developed. First, for each AB
pair, an illumination corrected A$+$B sky frame was produced as
described above and then subtracted from the parent A and B
frames. Second, on each frame, the standard procedure was
applied. This two-step procedure produced a much better background
correction than the standard procedure alone.

Each background corrected frame was then corrected for telluric
absorption lines, using the associated normalized telluric absorption
correction vector. As described in the cluster star processing section
above, small shifts in the dispersion direction as well as small
multiplicative scaling were necessary to achieve the best possible
correction.

Finally, the six (or eight) individual frames combined into a single
frame. Before this combination, the individual frames were shifted to a
common center (defined by the maximum intensities in the galaxy
spectra) with sub-pixel accuracy using linear interpolation. Given the
pixel scale (0\farcs148 pix$^{-1}$) vs. the natural seeing (estimated
to be $\sim$ 0\farcs8 from the profiles of the telluric star
observations), alignment accuracy is not critical. The frames were then
scaled to a common median value at a fixed point in the galaxy spectra
and then averaged using a sigma-clipping rejection scheme. The final
step was to multiple each combined frame by the sensitivity function
derived from its associated telluric standard (see above).

\begin{figure}
\resizebox{\hsize}{!}{\includegraphics[angle=0]{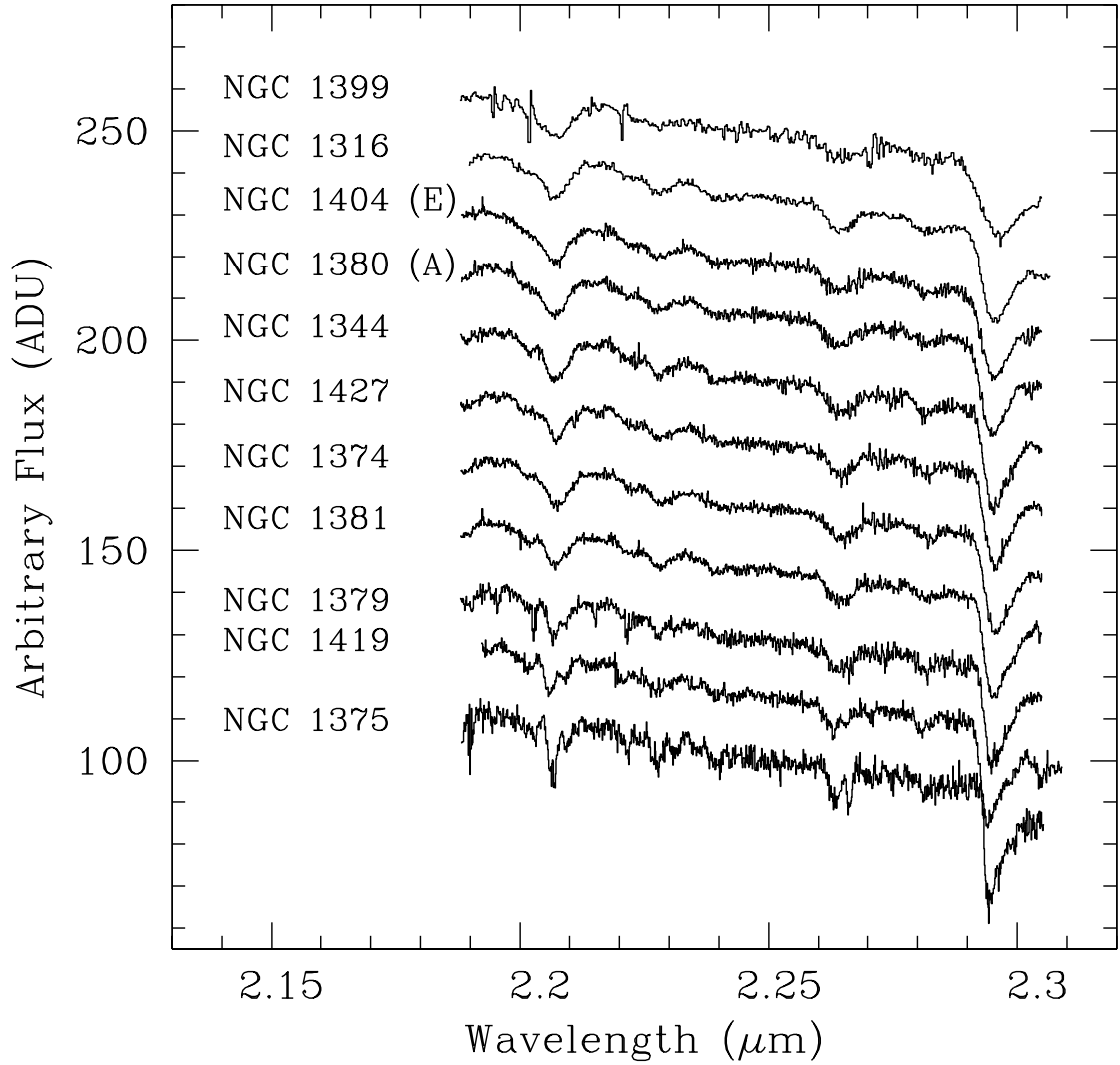}}
\caption{\label{fig:galCenters} Central 3\arcsec\/ diameter
  extractions for all galaxies ordered by decreasing central velocity
  dispersion from top to bottom (see Table~\ref{tab:galSample}).}
\end{figure}

Basic post-processing data quality can be assessed from
Figure~\ref{fig:galCenters} shows the central 3\arcsec\/ diameter
extraction for all the galaxies observed in 2002 November. The spectra
are ordered by velocity dispersion ($\sigma$) which corresponds
essentially to decreasing signal-to-noise per pixel due to decreasing
central intensity. The sharper spectral features seen in the galaxies
with lower $\sigma$ are intrinsic to the galaxies -- they are not
caused by less accurate telluric line correction in these fainter
galaxies.

%
%

\begin{figure}
\resizebox{\hsize}{!}{\includegraphics[angle=0]{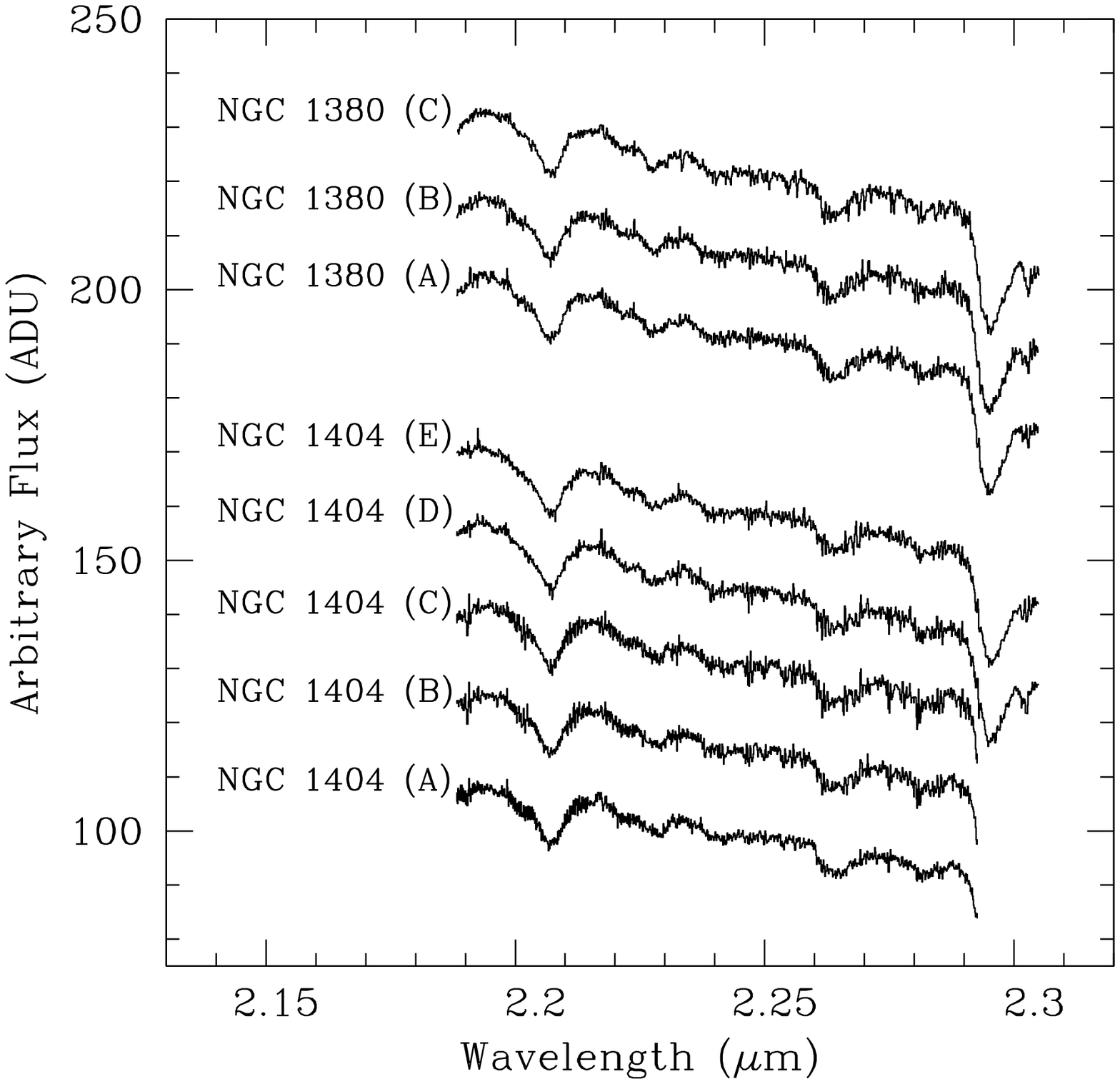}}
\caption{\label{fig:repeatCenters} Central 3\arcsec\/ diameter extractions for 1404(A-E) and
  1380(A-B).}
\end{figure}

To first order, measurement repeatability can be assessed from
Figure~\ref{fig:repeatCenters} where the central 3\arcsec\/ diameter
extraction of NGC\,1380 and NGC\,1404 are shown from two and three
different nights, respectively. In the case of NGC\,1404, these
observations were separated by months. Clearly, the gross properties
of the extracted spectra are the same. Repeatability of line strength
measurements is discussed in the next section.

The central spectra used for the measurement of spectroscopic features
were extracted within a radius corresponding to 1/8 R$_{e}$ at the
observed position angle for each galaxy (see also
Tables~\ref{tab:galSample} and \ref{tab:obsLog}).  For each extracted
spectra, an empirical signal-to-noise ratio (SNR) was derived
following Stoehr et al. (2007) and listed in
Table~\ref{tab:galSample}.

%
%

From these extracted spectra, recession velocities and velocity
dispersions were measured using the penalized Pixel-Fitting method
(pPXF) developed by Cappellari \& Emsellem (2004). For simplicity,
purely Gaussian line shapes were assumed. The resulting measurements
are needed as input for the line-strength determinations (see
Section~\ref{sec:lines}). The velocity dispersions are tabulated in
Table~\ref{tab:galSample} and are used throughput this paper. Errors
on the kinematics parameters are calculated within the code based on
error frames which are produced by our data reduction scripts. See
also Section~\ref{sec:proc-optical} for further analysis of the
kinematic errors.

\subsection{SINFONI data reduction}
\label{sec:proc-sinfoni}

SINFONI observations of NGC 1316 and NGC 1399 were processed and
calibrated using the SINFONI Pipeline (v.~1.6). The pipeline
transforms raw science data frames into flat-field corrected,
wavelength calibrated and background subtracted three-dimensional data
cubes for the galaxies and the calibration stars, including the
extraction of one-dimensional spectrum from the central 10 pixels of
the observed galaxies and stars.

For NGC\,1399, the telluric stars were solar-type. In order to obtain
a telluric correction spectrum, the star spectra were divided by a
scaled and Gaussian broadened solar spectrum. This step removes
intrinsic stellar features and continuum shape. A correction for
telluric features was then applied as described above for the ISAAC
observations.

The NGC\,1316 telluric stars were hotter B stars that have no
prominent spectral features in the wavelength range of interest.  The
telluric correction spectrum was obtained by dividing the stellar
spectrum by a scaled black body spectrum with the same temperature as
the star. A correction for telluric features was then applied as
described above for the ISAAC observations.

The individual data cubes for the two galaxies were then combined to
achieve better signal-to-noise ratios. The four NGC\,1399 data cubes
were averaged using a sigma-clipping pixel reject algorithm. Since
there were only two NGC\,1316 observations, pixel reject during
combination was not possible.


Since the field-of-view of the VLT-SINFONI observations was limited to
approximately 3$\times$3\arcsec\/, an extraction radius of 1/8 R$_e$
is not possible. Therefore, a slit of 2$\times$3\arcsec\/ was
extracted from the datacube to match the slitwidth of the optical data
for these galaxies (see Section~\ref{sec:proc-optical}). As for the
ISAAC data, the recession velocities and velocity dispersions were
measured with pPXF (see also Table~\ref{tab:galSample}) and errors
determined from noise statistics.

\subsection{Optical data processing}
\label{sec:proc-optical}

The long-slit spectroscopy from Kuntschner (2000) and Kuntschner et
al.  (2002, NGC\,1344) was used to extract a central spectrum for each
galaxy. The extraction aperture was set to match the effective area
covered on the sky by the central ISAAC data. The position angle of
the ISAAC observations was set to agree with the optical data. Due to
the rather different slit widths involved (ISAAC 1\arcsec, optical
Fornax data 2\farcs3 and 2\arcsec) the extraction ranges vary. For the
two galaxies observed with VLT-SINFONI (NGC\,1316 and NGC\,1399) the
extraction aperture is limited by the field-of-view of SINFONI. In
order to best match the optical and near-IR data we extracted a central
2\farcs3 $\times$ 3\arcsec slit from the optical observations.

Again pPXF was used to derive recession velocities and velocity
dispersions. Good fits were achieved for all galaxies but NGC\,1375
where the formal result was $\sigma = 32$\,\kms. This is well below
the resolution limit of the optical Fornax data (Kuntschner 2000;
4.1\,\AA\/ FWHM, R~$\simeq 1300$) and is therefore unreliable.  In
this paper, the velocity dispersion derived from the higher spectral
resolution ISAAC data was adopted and assigned a relatively large
error of $\pm$20\,\kms\/ to reflect the uncertainty of this procedure.

\begin{figure}
\resizebox{\hsize}{!}{\includegraphics{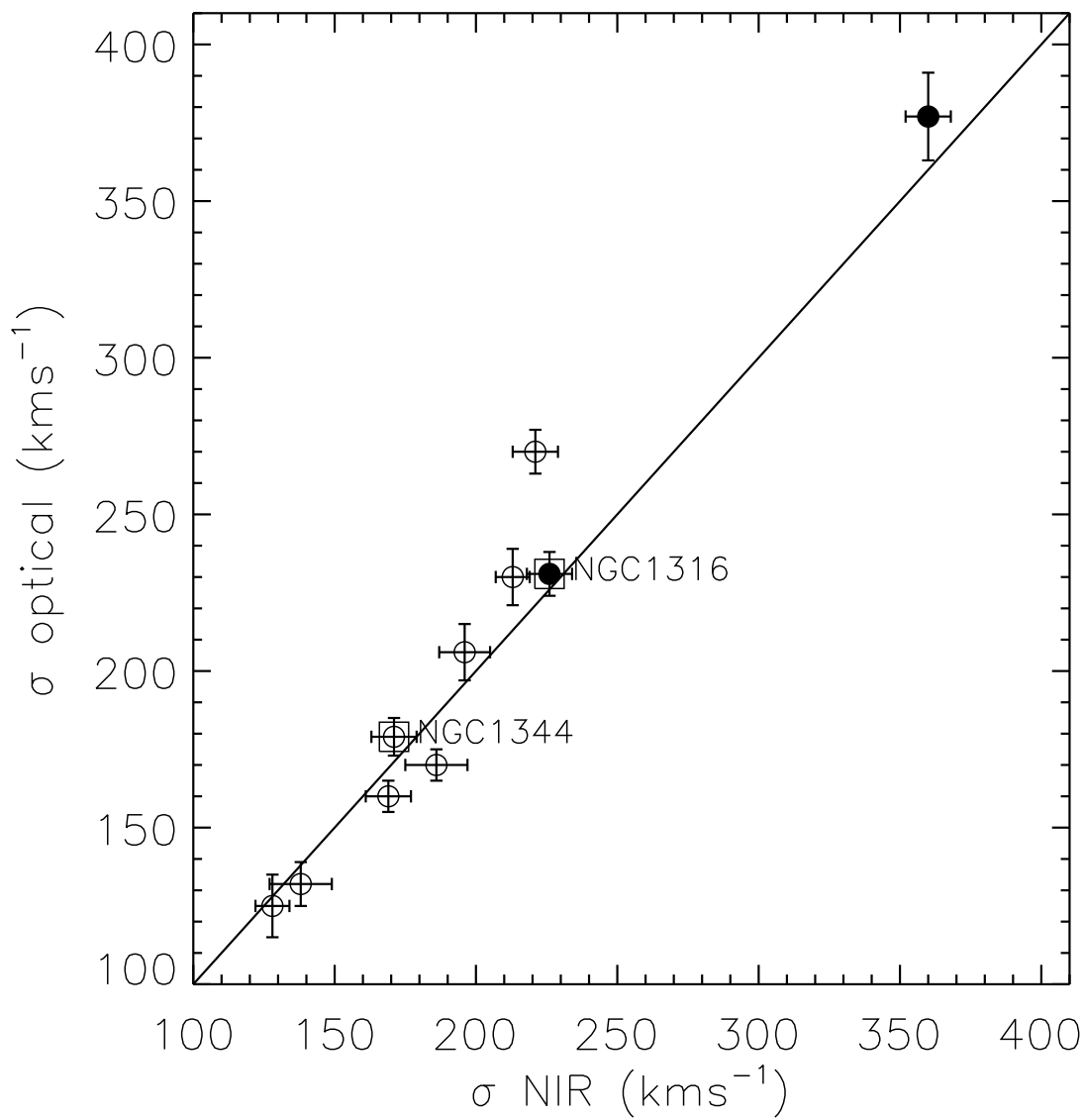}}
\caption{\label{fig:sigmas} Comparison of our velocity dispersions
  determinations between the near-IR data and the optical spectroscopy.
  Central extractions correspond to 1/8 R$_e$, except for NGC\,1316
  and NGC\,1399 ($\sigma \simeq 360$\,\kms) where a fixed aperture of
  2$\times$3\arcsec\/ was used. NGC 1375 is not shown -- see text for
  details.}
\end{figure}

A comparison between optical and near-IR measurements of the central
velocity dispersions is shown in Figure~\ref{fig:sigmas}. The
determinations are in good agreement with each other for all galaxies,
including the two galaxies with significant contributions of young
stellar populations (NGC\,1344 and NGC\,1316). Taking the difference
between the optical and near-IR measurements as a guideline for our
true errors (including differences in seeing and remaining aperture
differences) we conclude that our internal velocity dispersion error
needs to be scaled by a factor of 2.2. In the following analysis and
figures, these conservative scaled errors are used.

%
%

\section{Line Strength Measurements}
\label{sec:lines}
\subsection{Spectral Feature Definitions}
The spectral features investigated are listed in
Table~\ref{tab:features}. For all but one feature, absorption-line
strength is measured as equivalent width via indices, where a central
feature bandpass is flanked to the blue and red by pseudo-continuum
band-passes. This method is identical to what is used to measure
optical Lick indices (Worthey et al. 1994). The index definitions for
\naK\/ and \caK\/ were taken from Frogel et al.  (2001). Revised \fea,
\feb, and \mg\/ index definitions were used since earlier definitions
(e.g. F{\"o}rster-Schreiber 2000) did not incorporate continuum band
definitions optimized for galaxy spectra with significant velocity
broadening. In order to measure the strength of the $^{12}$CO(2,0)
absorption we use the definitions of Frogel et al.  2001 which provide
four pseudo-continuum bandpasses to the blue of the CO feature and
none to the red. Therefore the continuum used to derive the equivalent
width is an extrapolation from the blue wavelength region in contrast
to the Lick indices where the continuum is based on an interpolation
between two flanking bandpasses.

\begin{figure}
\resizebox{\hsize}{!}{\includegraphics[angle=0]{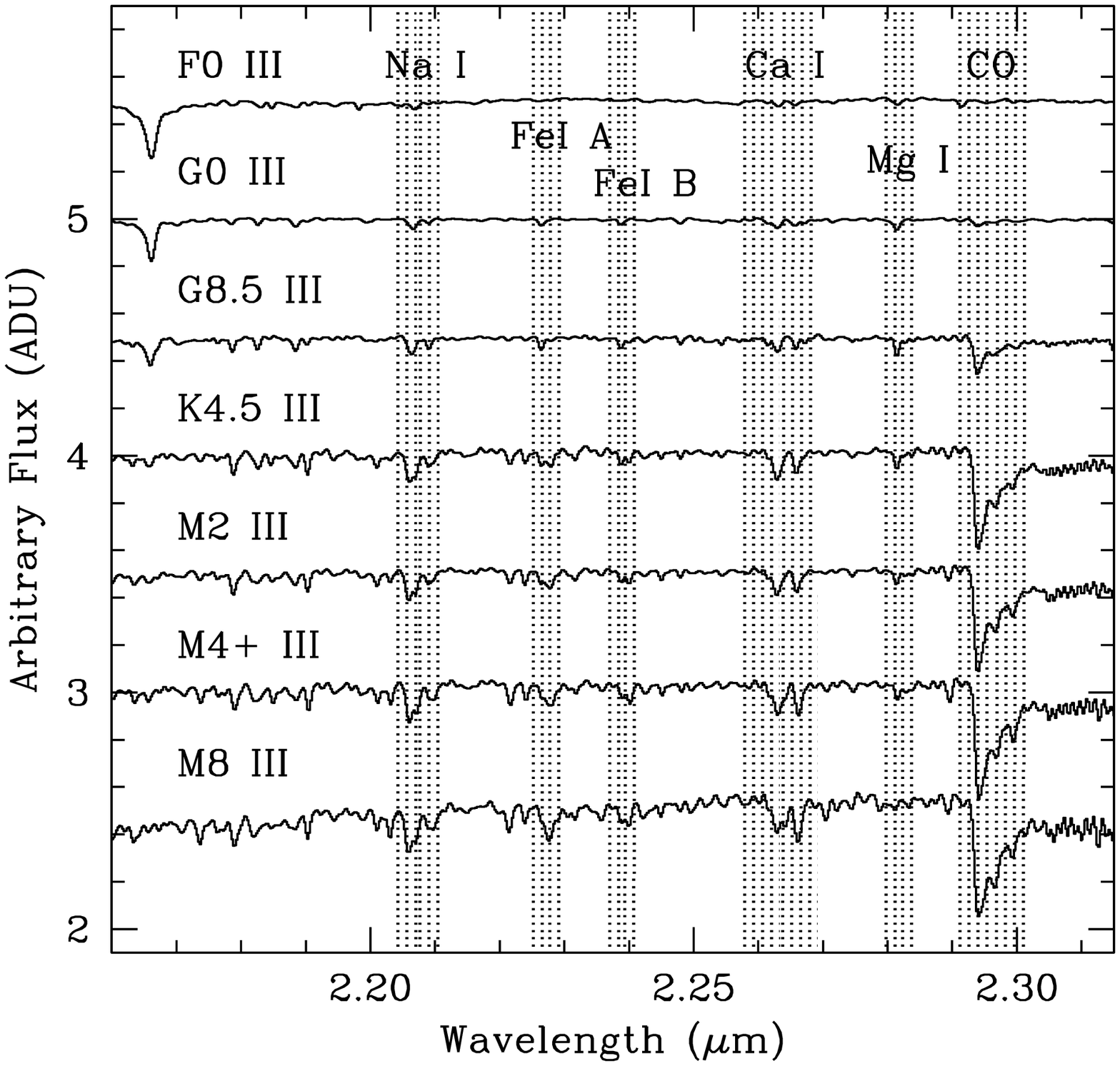}}
\caption{\label{fig:features} Spectral Features Defined -- Using
near-IR spectra from Wallace \& Hinkle (1997), the on-band regions of
the spectral features defined in Table~\ref{tab:features} are
shown. The Wallace \& Hinkle spectra and the new spectra described
here have similar spectral resolution (see
Figure~\ref{fig:clusterStars}).}
\end{figure}

The feature bandpasses of all measured indices are illustrated in
Figure~\ref{fig:features} which shows how the absorption features
change as a function of temperature in red giant stars. The spectra
shown (taken from Wallace \& Hinkle 1997) are similar in spectral
resolution to our spectra (see Figure~\ref{fig:clusterStars}
above). It is important to remember that all of these features are
really blends of distinct atomic and molecular features, especially at
cooler temperatures. This point is richly illustrated by the R $\sim$
40 000 spectra presented in Wallace \& Hinkle (1996) (see, in
particular, their Figures 2 and 3 for the \caK\/ and \naK\/ features,
respectively).

%
%
\begin{deluxetable}{lccccl}
\rotate
\tablewidth{0pt}
\tablecolumns{6}
\tabletypesize{\small}
\tablecaption{\label{tab:features} Index definitions}

\tablehead{
\colhead{Index}
&\colhead{Blue pseudo-continuum}
&\colhead{Central bandpass}
&\colhead{Red pseudo-continuum}
&\colhead{Units}
&\colhead{Source}
}

\startdata
Na\,{\small I}    & 21910 -- 21966 & 22040 -- 22107 & 22125 -- 22170  & \AA &   Frogel et al. 2001\\
Fe\,{\small I}\,A & 22133 -- 22176 & 22250 -- 22299 & 22437 -- 22497  & \AA &   This paper$^a$\\
Fe\,{\small I}\,B & 22133 -- 22176 & 22368 -- 22414 & 22437 -- 22497  & \AA &   This paper$^a$\\
Ca\,{\small I}    & 22450 -- 22560 & 22577 -- 22692 & 22700 -- 22720  & \AA &   Frogel et al. 2001\\
Mg\,{\small I}    & 22700 -- 22720 & 22795 -- 22845 & 22850 -- 22874  & \AA &   This paper$^a$\\
\tableline
Index          & \multicolumn{2}{c}{Continuum bands} & Feature bandpass & Units & Source \\
$^{12}$CO(2,0) & \multicolumn{2}{c}{22300 -- 22370, 22420 -- 22580, 22680 -- 22790, 22840 -- 22910} & 22910 -- 23020 & \AA & Frogel et al. 2001 \\
\enddata

\tablecomments{Notes: (a) Our new index definitions were inspired by
F{\"o}rster-Schreiber (2000) but are not identical to the definitions
used in that paper.}
\end{deluxetable}

%
%

\subsection{Measurement Procedure}
In order to allow a meaningful comparison between observations of
stars and galaxies (with different internal dynamics) to themselves
and to population models, index measurements need to be carefully
calibrated to a common system. There are generally three effects to
account for: (a) the differences in spectral resolution of the
instruments used to obtain the index measurements; (b) the internal
velocity broadening of galaxies and (c) possibly small systematic
offsets caused by continuum shape differences (see also Kuntschner
2000).

The resolution of our ISAAC observations (6.9\AA\/ FWHM, or R~$\simeq
3300$) was adopted as the nominal resolution. The VLT-SINFONI spectra
were broadened to match this resolution before the measurement of the
line-strengths indices.  The observed spectrum of a galaxy is the
convolution of the integrated spectrum of its stellar population(s) by
the instrumental broadening and line-of-sight velocity distribution
(LOSVD) of the stars. These effects typically broaden the spectral
features, in general reducing the observed line-strength compared to
the intrinsic values. In order to compare the index measurements for
galaxies we calibrate the indices to zero velocity dispersion and our
nominal resolution.

Although non-Gaussian deviations from the LOSVD can have significant
effects on the LOSVD correction of indices (see e.g., Kuntschner
2004), in this study, only the first moments, $v$ and $\sigma$, are
considered. This is a reasonable approximation for central spectra
studied in this study since expected non-Gaussian deviations in this
region are negligible.

Observations of K and M giants are used to determine the corrections
for each index.  These spectra are a reasonable match to the observed
galaxy spectra; however, once stellar population models become
available the corrections should be re-derived. By broadening the
stellar spectra to velocity dispersions of up to 400\,\kms\/ in steps
of 20\,\kms\/ a median correction factor is computed for each index,
i.e.:

\begin{equation}
C_{j}(\sigma) = I_{j}(\sigma=0) / I_{j}(\sigma),
\end{equation}

\begin{figure}
\resizebox{\hsize}{!}{\includegraphics[angle=0]{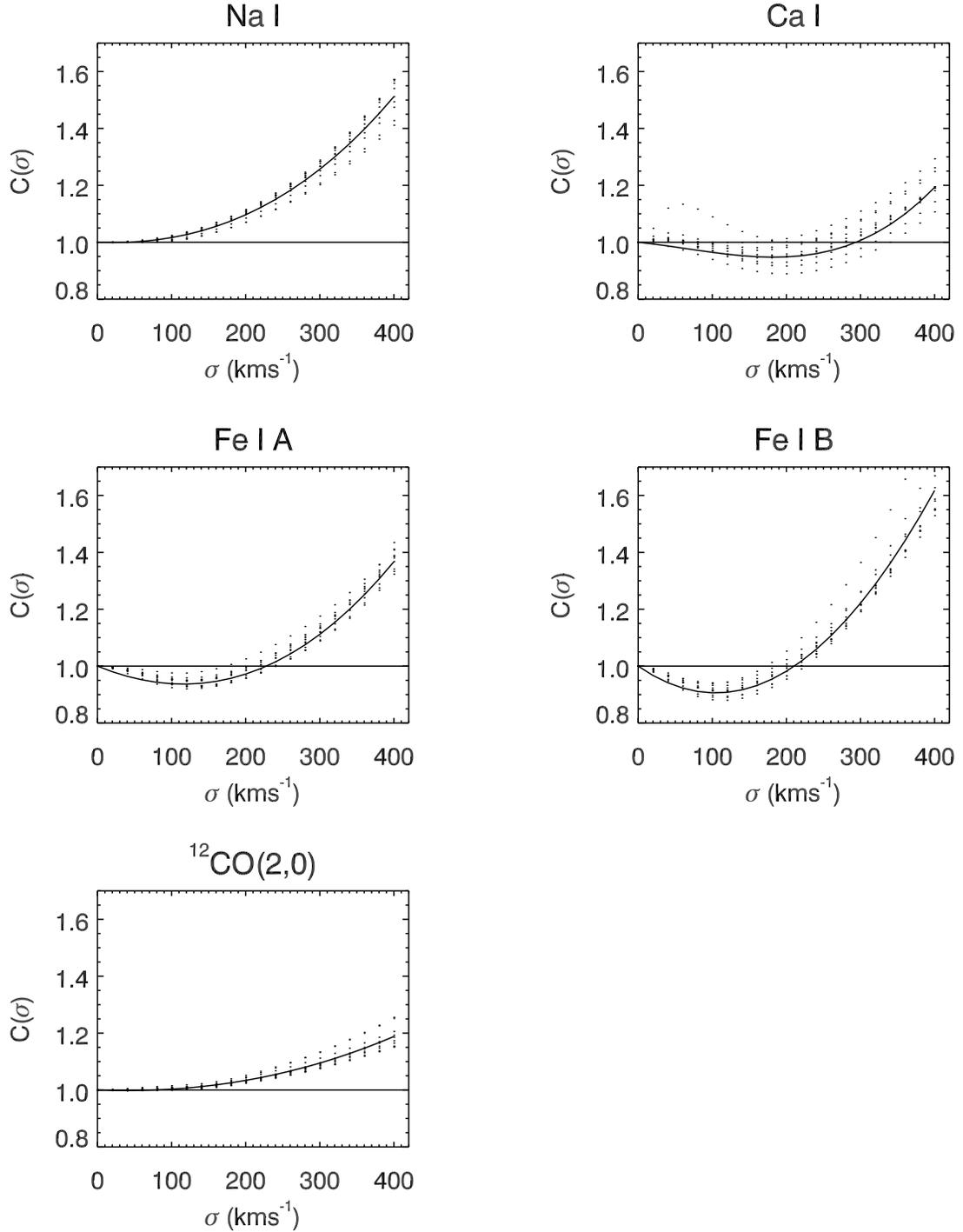}}
\caption{\label{fig:veldispcorr} Line of sight velocity distribution
(LOSVD) corrections for near-IR indices. Dots represent individual
stars broadened to different velocity dispersions. The solid line is a
3rd order polynomial fit to the data and is used to apply the LOSVD
corrections.  See text for further details.}
\end{figure}

\par\noindent and $I_{j}$ is the median index measured from the stars
convolved with the LOSVD given in brackets. A LOSVD corrected index is
then $I_{j}^{corr} = C_{j}(\sigma) \times I_{j}^{raw}$. The actual
correction functions computed here are shown in
Figure~\ref{fig:veldispcorr}.

%
%

For the optical data the velocity dispersion corrections from
Kuntschner (2004) were used. The errors on the line-strength indices
are evaluated by using error estimates from the spectra and taking into
account the errors from the LOSVD correction procedure. The optical
data was calibrated to the Lick/IDS system. For a more detailed
description of the procedure see Kuntschner (2000).

\begin{figure}
\resizebox{\hsize}{!}{\includegraphics[angle=0]{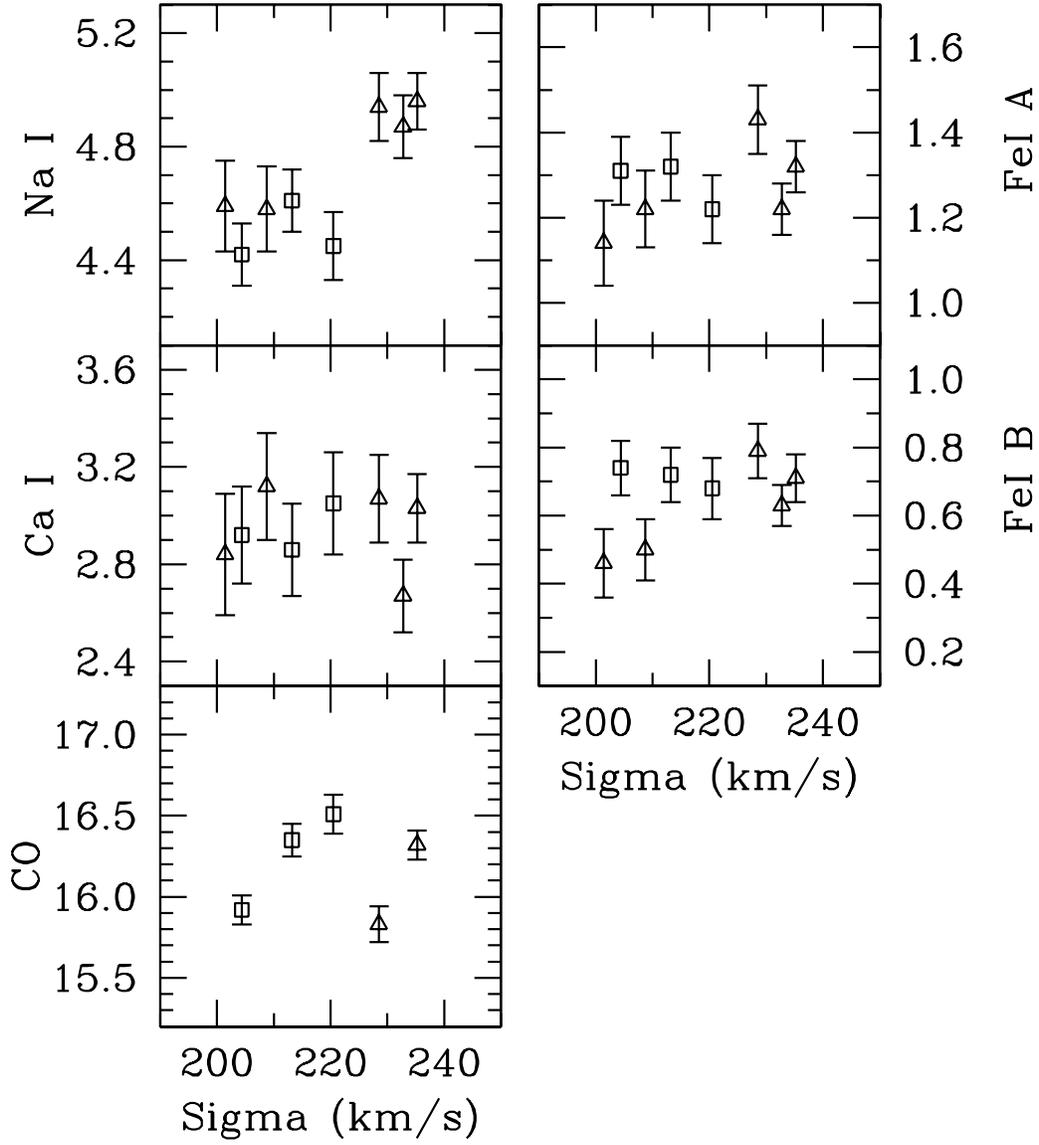}}
\caption{\label{fig:repeatability} Repeated ISAAC Line Strength
Measurements. Repeated measurements for NGC 1380 ({\it open squares})
and NGC 1404 ({\it open triangles}) are shown. See text for further
information.}
\end{figure}

Multiple ISAAC observations of NGC\,1380 and NGC\,1404 allow us to
constrain possible systematic errors in our line strength
measurements.  Figure~\ref{fig:repeatability} is plotted with the same
y-axis range as Figure~\ref{fig:galLineSigma} but with an smaller
$\sigma$ range.  For the \naK, \caK, \fea\/ and \feb\/ indices our
error estimate are comparable to the observed repeatability
uncertainty. However, in the case of the \co\/ index the repeatability
uncertainty is significantly larger and we adopt the internal
dispersion of the NGC 1380 and NGC 1404 measurements (0.3\,\AA) as
the formal uncertainty for all subsequent analysis.

%
%

Line-strength indices can be sensitive to continuum shape calibration
errors. In particular index definitions spanning a wide wavelength
range are prone to suffer from this. In our study the wide band-pass of
the CO index (see Table~\ref{tab:features}) falls in this regime and we
consider this to be the explanation for the larger spread between
repeat observations. 


\section{Results}
\label{sec:results}

\subsection{Cluster stars}

The origin and behavior of the \naK, \caK, and \co\/ features as a
function of effective temperature in K and M giant stars has been
extensively reviewed and discussed by Ram\'{i}rez et al. (1997) and
F\"{o}rster Schreiber (2000) (who also studied the Fe and Mg
features).  Most of the stars used in those studies were field stars
with bright apparent magnitudes and hence drawn from a narrow
metallicity range biased towards solar metallicity. The behavior of
the \naK, \caK, and \co\/ features as a function of metallicity for
Galactic globular cluster K and M giants has been investigated by
Frogel et al. (2001). The metallicity range of that study covers both
the metallicity range of the clusters investigated here, as well as
the luminosity-weighted mean metallicities of the Fornax early-type
galaxies.

\begin{figure}
\resizebox{\hsize}{!}{\includegraphics[angle=0]{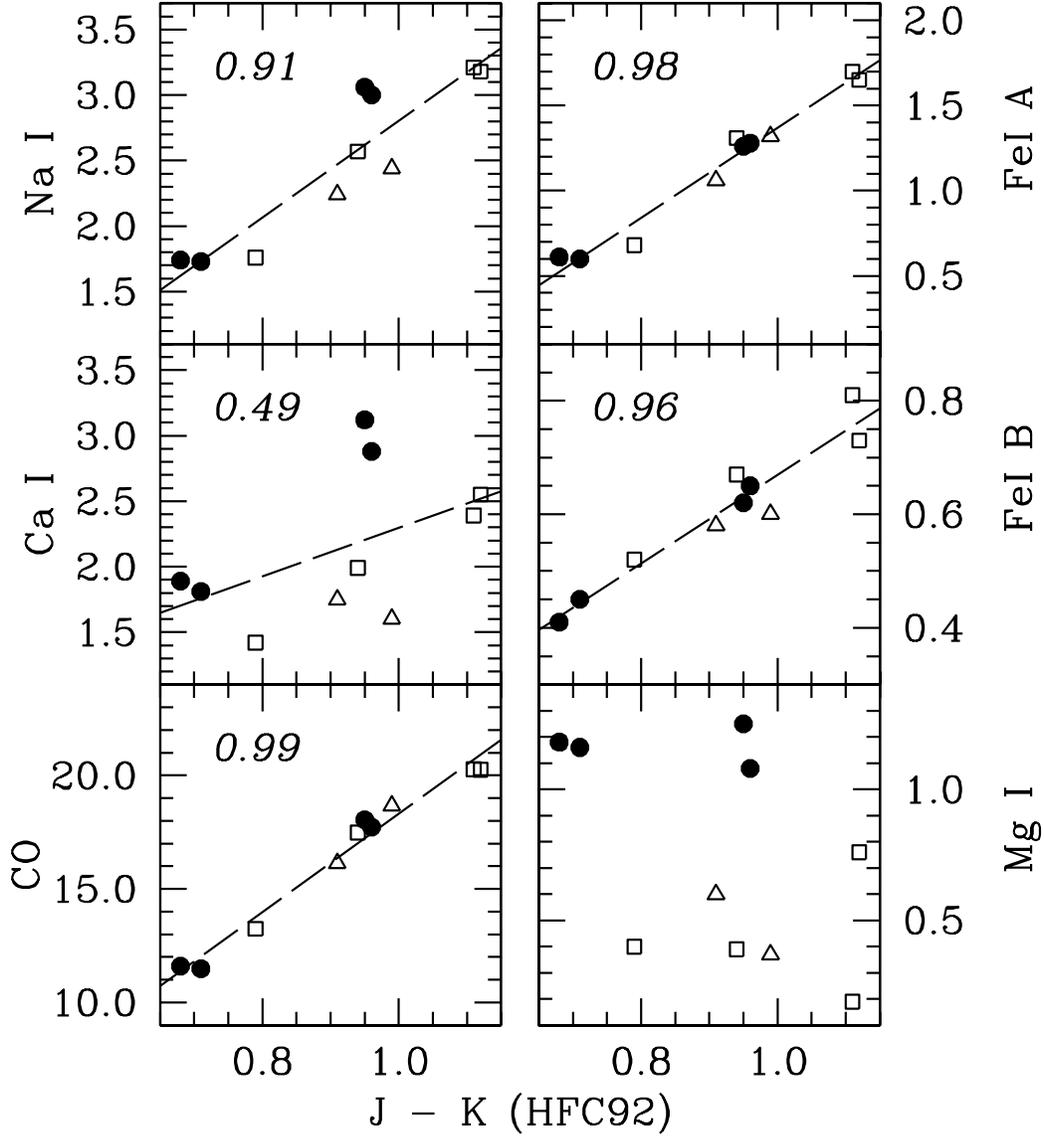}}
\caption{\label{fig:starLineColor} Cluster Star Line Strength
vs. $J-K$. The strengths of all six measured K-band spectral features
plotted against $J-K$ color as tabulated by Houdashelt, Frogel, \&
Cohen (1992). The {\it filled circles} are NGC 2477 stars ([Fe/H] $=$
--0.02), the {\it open squares} are NGC 2204 stars ([Fe/H] $=$
--0.38), and the {\it open triangles} are NGC 2506 stars ([Fe/H] $=$
--0.52). The typical random uncertainty is smaller than the symbols,
while the typical HFC92 color uncertainty is 0.02 mag. The {\it dashed
line} in each panel is the unweighted linear least-squares fit between
index strength and color. Linear correlation coefficients are given in
each panel. \mg\/ appears to be constant within a cluster (or rather
metallicity); hence, no global trend with color has been derived.}
\end{figure}

In Figure~\ref{fig:starLineColor}, the index strength (in \AA) for all
six measured spectral features is plotted against $J-K$ color as
tabulated by Houdashelt, Frogel, \& Cohen (1992) (hereafter HFC92)
for the open cluster stars listed in Table~\ref{tab:starSample}. These
stars are all K or M giant stars. In this context, redder $J-K$
implies cooler effective temperature and lower surface gravity. Within
a single cluster, it is reasonable to assert that this lower effective
temperature is not metallicity-driven. As previously shown in the
studies discussed above, all features increase in depth with redder
color (and hence decreasing $T_{eff}$) except \mg, which appears to
have a more or less constant value for each cluster regardless of
effective temperature (see also F\"{o}rster Schreiber 2000).

\begin{figure}
\resizebox{\hsize}{!}{\includegraphics[angle=0]{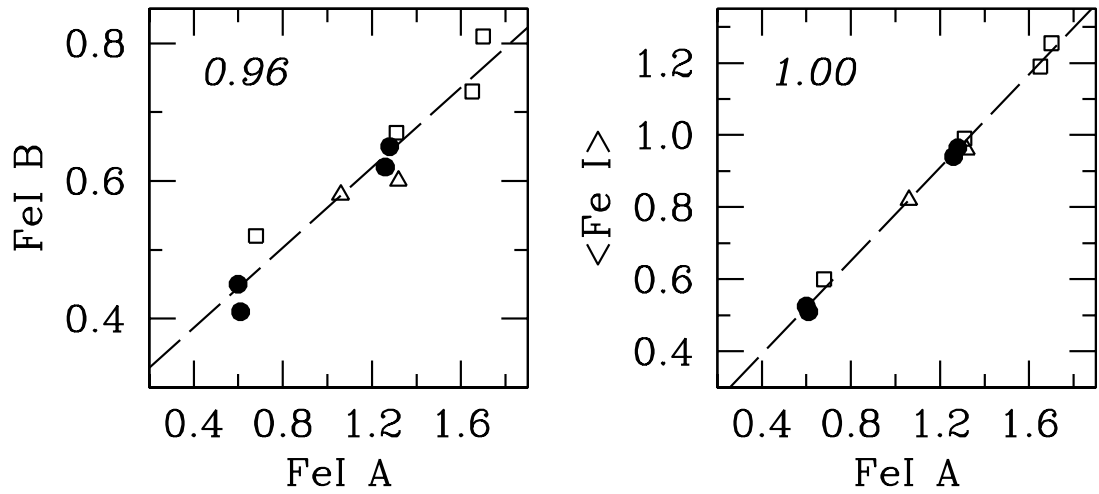}}
\caption{\label{fig:starFe} \fe\/ Definition. Since FeI A and FeI B
  strengths are highly correlated ({\it left panel}), it is convenient
  to define a mean index \fe\/ = (FeI A + FeI B) / 2.  The symbols and
  dashed line are explained in Figure~\ref{fig:starLineColor}. The
  linear correlation coefficient for each panel is written in the
  upper left corner.}
\end{figure}

The \co, \fea, and \feb\/ features are highly correlated with $J-K$
and show no dependency on cluster metallicity. For the CO feature,
this behavior is consistent with the results of Frogel et al. (2001)
over a similar cluster metallicity range. No analogous study exists
for the Fe features. However, as shown in Figure~\ref{fig:starFe}, the
\fea\/ and \feb\/ index strengths are highly correlated with each
other. Hence, it is convenient to define a mean index: \fe\/ = (FeI A
+ FeI B) / 2 that has a smaller formal uncertainty than either
individual Fe index alone.  This mean index \fe\/ is used throughout
the rest of this paper.

As anticipated by the results of Frogel et al. (2001), at a given
$J-K$ (i.e. $T_{eff}$), giant stars in more metal-rich clusters have
stronger \naK\/ and \caK\/ features. The Mg feature (not included in
the Frogel et al. study) appears to have similar behavior, although we
note again that within a given cluster (and hence for a given
metallicity), the Mg feature strength is approximately constant with
$J-K$.

\begin{figure}
\resizebox{\hsize}{!}{\includegraphics[angle=0]{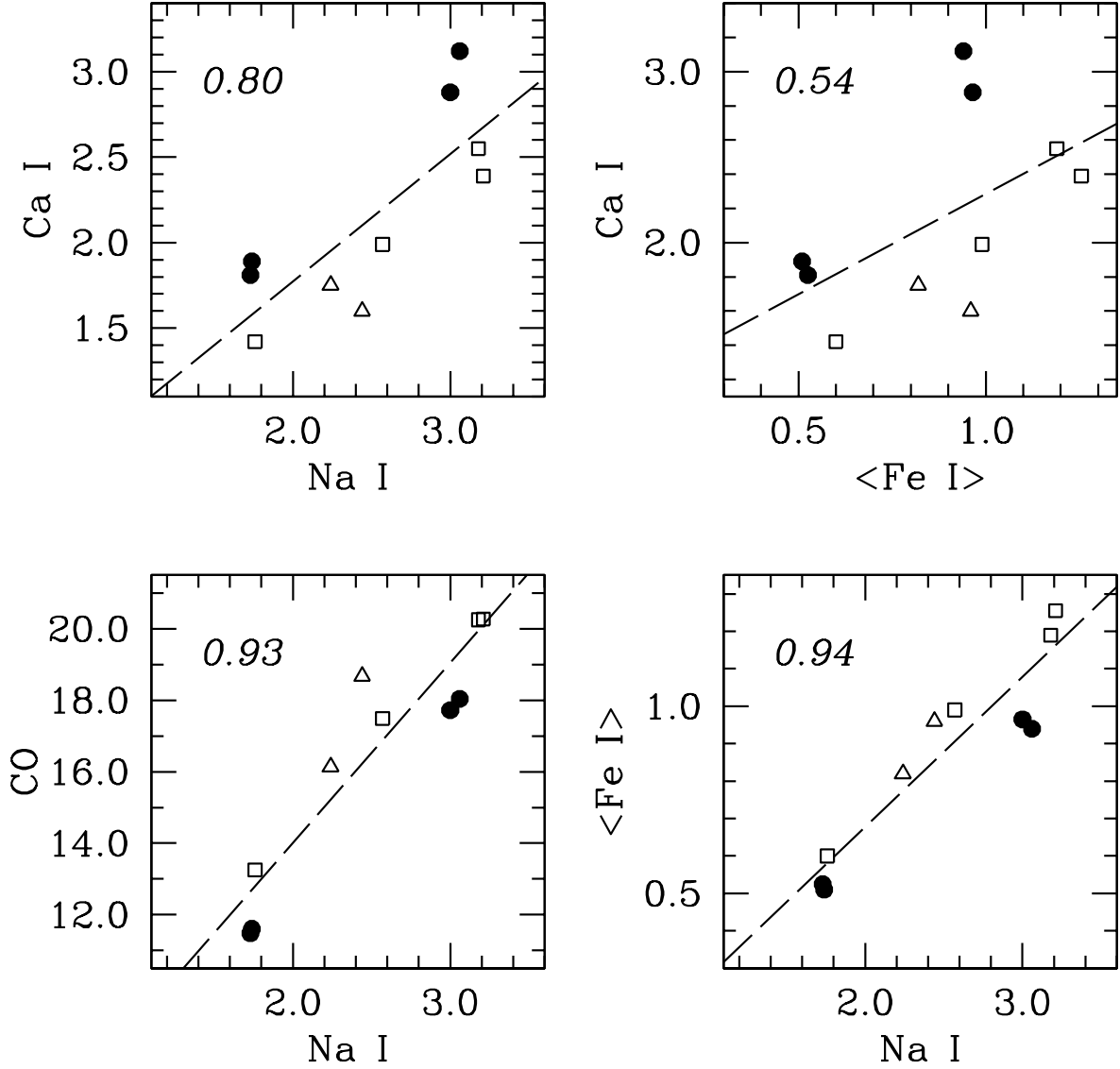}}
\caption{\label{fig:starLineLine} Cluster Star Line Strength vs. Line
  Strength. The symbols are explained in
  Figure~\ref{fig:starLineColor}. The dashed lines are the unweighted
  linear regression fits. The linear correlation coefficient for each
  relationship given in the upper left corner of each panel.}
\end{figure}

In Figure~\ref{fig:starLineLine}, various combinations of K-band line
indices are shown.  \naK, \fe\, and \co\/ are highly correlated with
each other (linear correlation coefficient, $r>$ 0.9). There is a hint
that \naK\/ is stronger in the solar-metallicity stars at equivalent
\co\/ and \fe\/ than in the more metal-poor stars. It is much clearer
that at equivalent \naK\/ and \fe, \caK\/ is stronger in the
solar-metallicity stars than in the more metal-poor stars.

%
%

%
%

%
%

\subsection{Fornax galaxies}
\label{sec:ls_fornax}

%
%
\begin{figure}
\resizebox{\hsize}{!}{\includegraphics[angle=0]{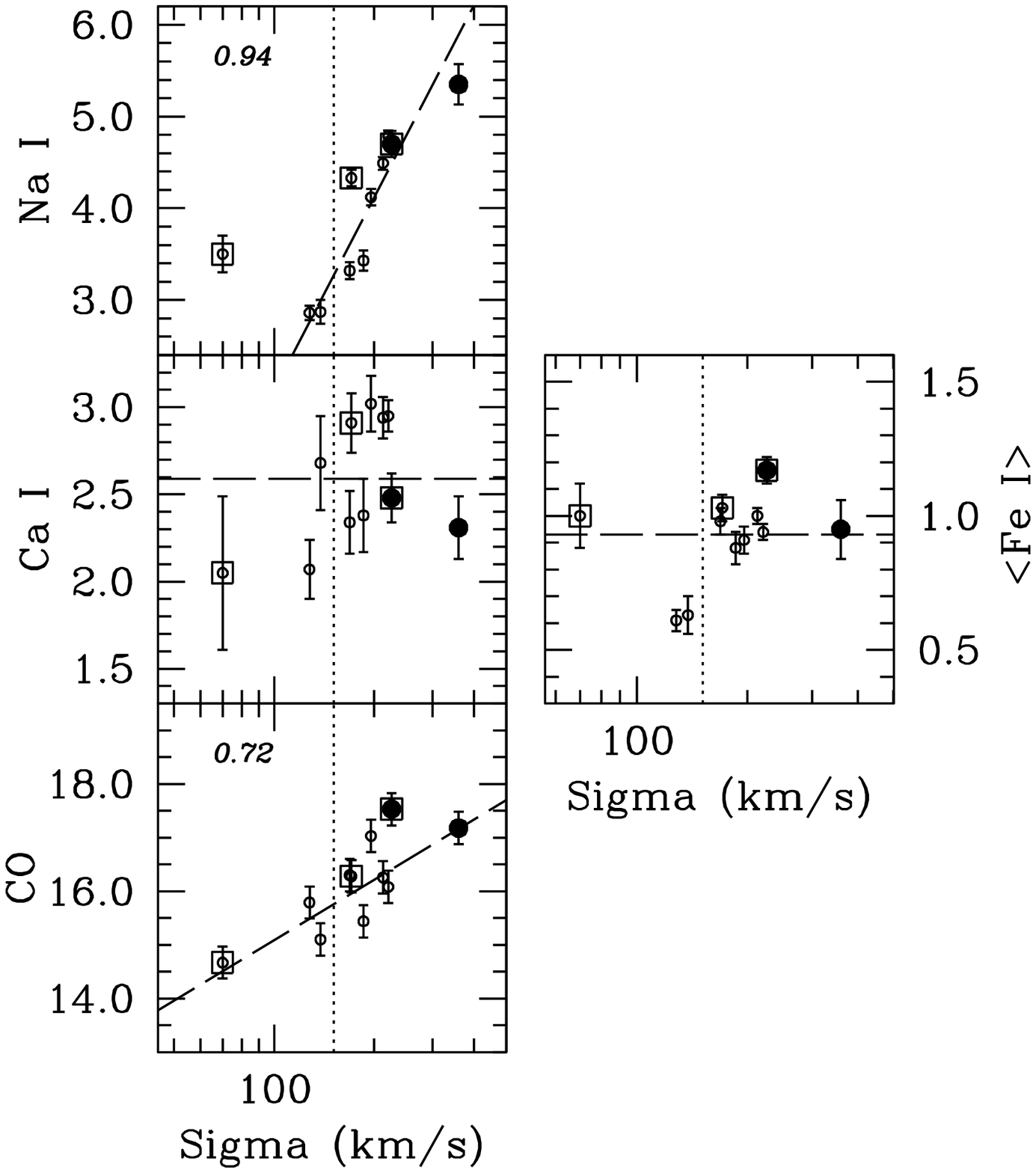}}
\caption{\label{fig:galLineSigma} Fornax E/S0 Line Strength
vs. Central Velocity Dispersion - the index strength plotted against
central velocity dispersion ($\sigma$) as measured from our K-band
spectra. Individual ISAAC measurements for NGC 1380 and NGC 1404 have
been averaged into mean values. The galaxies with young stellar
components (NGC 1316, NGC 1344, and NGC 1375) are marked with {\it
open squares}. Values derived from SINFONI observations (NGC 1316 and
NGC 1399) are indicated by {\it filled circles}.  The {\it slanted
dashed lines} in \naK, \caK, and \co\/ panels are the unweighted
linear regression fits for the purely old galaxies (i.e. without NGC
1316, NGC 1344, or NGC 1375).  The linear correlation coefficient for
each fit is shown. The {\it vertical dashed line} indicates $\sigma =
$150 \kms. In the \caK\/ and \fe\/ panels, the {\it horizontal dashed
lines} are the unweighted mean values for old galaxies with $\sigma >
150$ \kms.}
\end{figure}

In Figure~\ref{fig:galLineSigma}, central K-band spectral feature
measurements are plotted against central velocity dispersion measured
from our data.  \mg\ has not been shown here (and is ignored for the
rest of the paper) because this line is relatively weak and a reliable
velocity dispersion correction could not be derived. Relative to the
stellar spectral line strengths shown in
Figure~\ref{fig:starLineColor}, the galaxy measurements (except \naK)
fall in the middle of the ranges of the stellar measurements. Taken at
face value, this is consistent with saying that the
luminosity-weighted spectral type for all these galaxies is similar to
a solar-metallicity M0 III star.

For most of our galaxies, the \naK\/ feature is stronger than measured
in any of our observed stars or any of the field disk giants observed
by Ramirez et al. (1997). Only the supergiants in the F\"{o}ster
Schreiber have such strong Na features. The unlikely possibility of a
supergiant-dominated population can be ruled out by \co\/ index values
that are completely compatible with first-ascent giant stars. The
\co\/ feature is too strong for these galaxies to be dwarf dominated.

At equivalent $\sigma$, the galaxies containing younger populations
are separated from purely old galaxies. In the galaxies with purely
old populations, \naK\/ is highly correlated with $\sigma$, while
\co\/ and $\sigma$ are less well-correlated.  In contrast, neither
\caK\/ nor \fe\/ are well-correlated with $\sigma$ -- \fe\/ reaches a
maximum value for $\sigma \gtrsim 150$ while \caK\/ has significant
scatter at any given $\sigma \gtrsim 150$.\footnote{\caK\/ is
well-correlated with $\sigma$ and \mgfe\/ in purely old galaxies if
NGC 1399 is removed from consideration. However, there is no sound
reason for such data filtering at this time. Obviously, a larger
sample is needed to investigate this issue.}

%
%
\begin{figure}
\resizebox{\hsize}{!}{\includegraphics[angle=0]{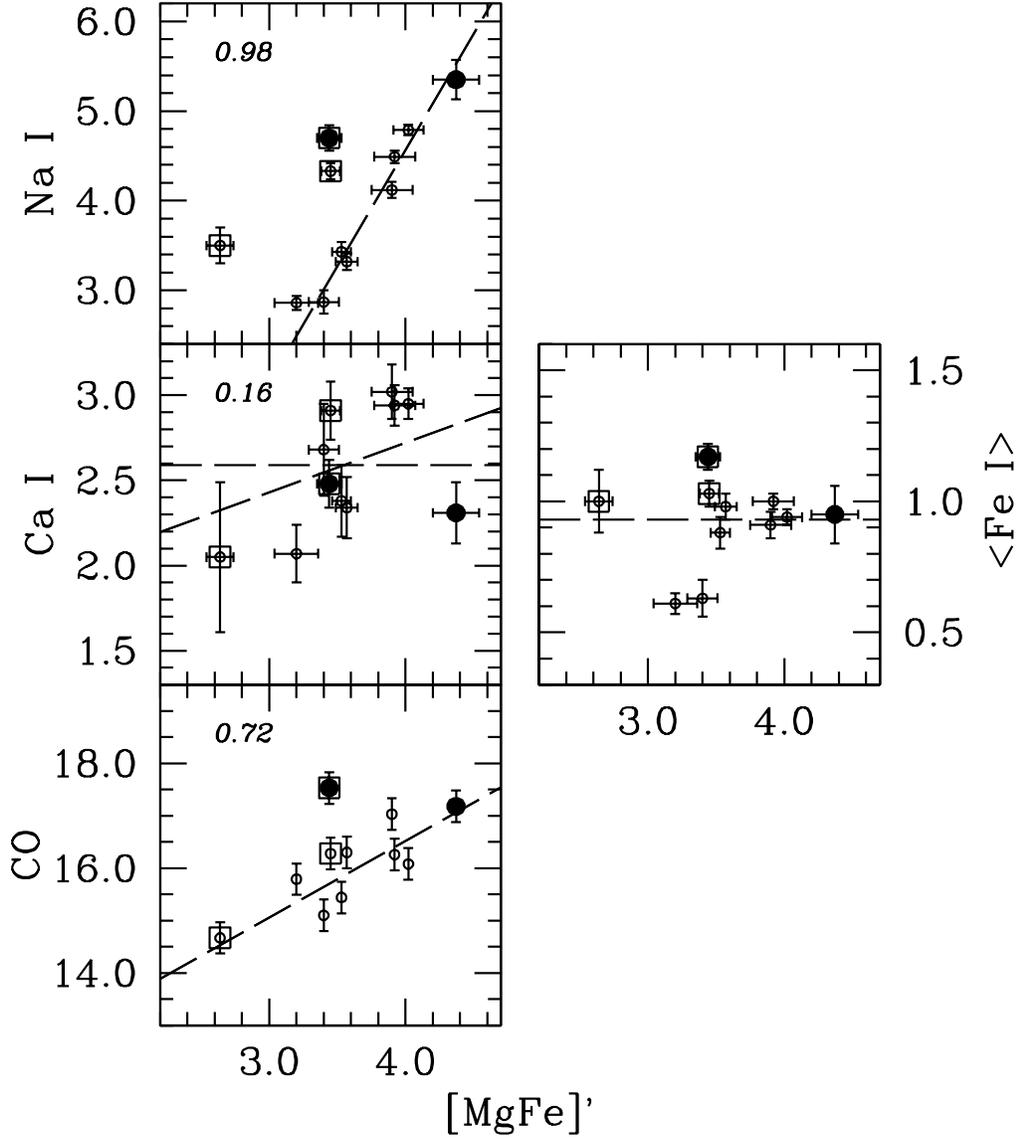}}
\caption{\label{fig:galLineMgFePrime} K-band feature strength
vs. \mgfe. \mgfe\/ is an optical index combination sensitive to total
metallicity \citep{tho03}. See Figure~\ref{fig:galLineSigma} for an
explanation of symbol type and dashed line.}
\end{figure}

Figure~\ref{fig:galLineMgFePrime} shows the relationship between the
optical index \mgfe\/ and our new K-band index measurements.  \mgfe\/
is an abundance ratio insensitive optical index combination in the
Lick system, designed to track total metallicity \citep{tho03}.  It is
known to be well correlated with $\sigma$; hence, it is not surprising
that Figure~\ref{fig:galLineSigma} and
Figure~\ref{fig:galLineMgFePrime} look very similar for the purely old
galaxies. Again, the galaxies that contain a young population are
separated from the purely old galaxies at equivalent \mgfe\/ and
$\sigma$.

%
%
\begin{figure}
\resizebox{\hsize}{!}{\includegraphics[angle=0]{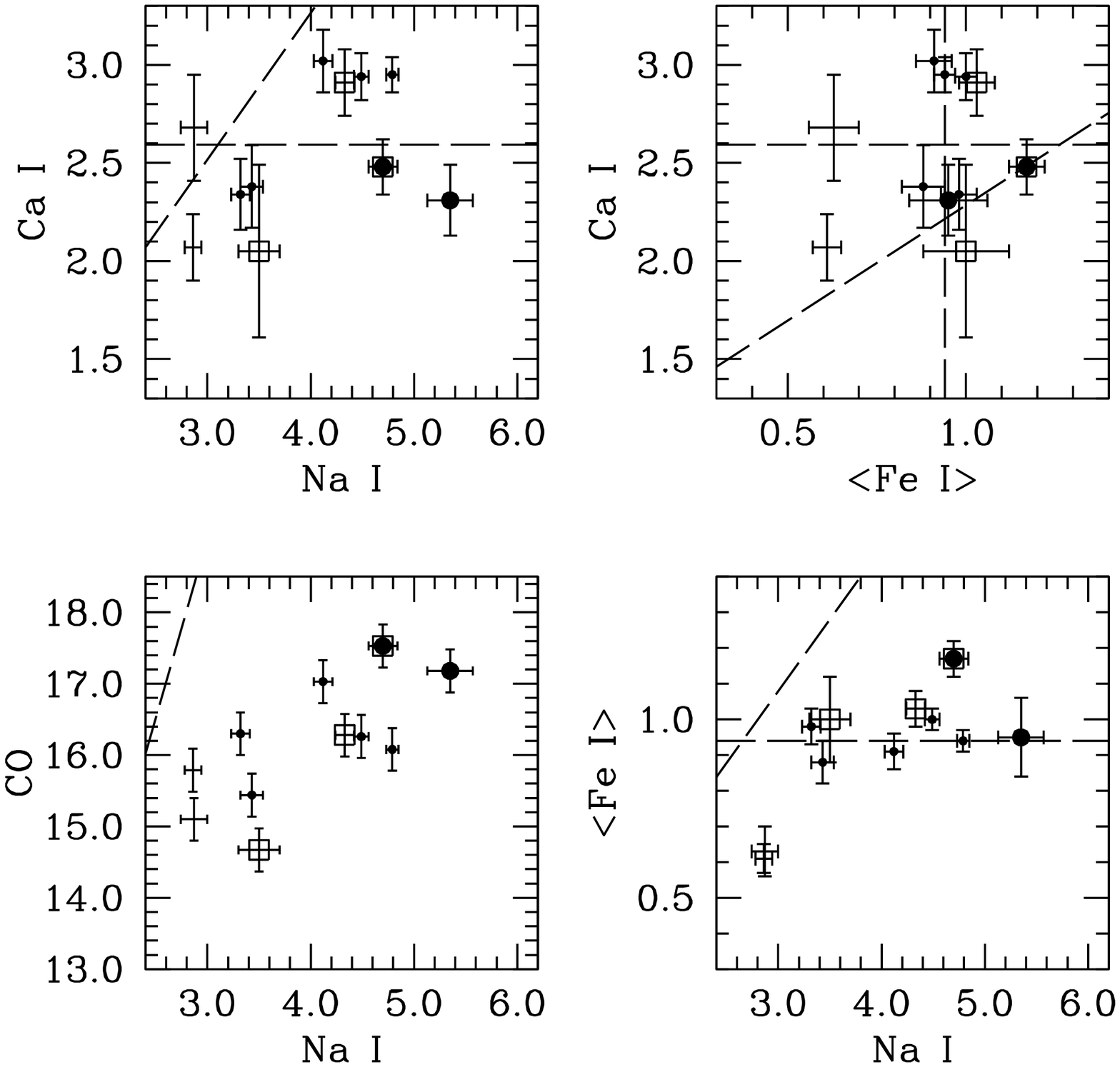}}
\caption{\label{fig:galLineLine} K-band index vs. index
relationships. See Figure~\ref{fig:galLineSigma} for a symbol
explanation. The {\it leftmost dashed line} in each figure are the
cluster star fits shown in Figure~\ref{fig:starLineLine}. The {\it
horizontal dashed lines} in the \caK\/ and \fe\/ panels are the mean
values for purely old galaxies with $\sigma > 150$ \kms. The {\it
vertical dashed line} in the \caK\ vs.~\fe\/ panel is the mean \fe\/
value.}
\end{figure}

Figure~\ref{fig:galLineLine} shows the relationships between various
pairs of K-band spectral indices (compare to
Figure~\ref{fig:starLineLine}). This figure captures several important
observational conclusions already foreshadowed above:

\begin{itemize}

\item{} Relative to the observed cluster stars, the measured values of
\naK\/ are significantly stronger in the galaxies. 

\item{} In all cases, the galaxies that contain signatures of younger
populations (i.e. stronger \hb\/ and weaker \mgfe\/) have stronger
\naK\/ and Fe features than comparable purely old galaxies.

\item{} \fe\/ saturates circa $\sigma \sim$ 150 \kms\/ (\naK\/ $\sim$
3.2) in the same galaxies. 

\end{itemize}

\section{Discussion}
\label{sec:disc}

No self-consistent theoretical spectral synthesis models for the
interpretation of integrated K-band spectra of early-type galaxies are
widely available. Development of such models is on-going (e.g. Marmol
Queralt, 2007, in preparation). Until such models exist, we must rely
on basic (and perhaps imperfect) astrophysical intuition and available
tools, such as the high-resolution near-IR spectral atlas of Wallace
\& Hinkle (1996) (hereafter WK96).

\subsection{Galaxies with young populations}

Three galaxies in our sample (NGC 1316, NGC 1344, and NGC 1375) have
stronger \hb\/ and weaker \mgfe\/ features than the rest of our sample
(see Figure~\ref{fig:fornax_pops}). This suggests a simple, two
component stellar population model. One stellar component is cold --
in the luminosity-weighted mean, it is presumably old ($>$ 8 Gyr) and
metal-rich ([Fe/H] $>$ --0.3) with spectral properties consistent with
the observed central velocity dispersion. In other words, it has
properties similar to the purely old galaxies in our sample (see
Figure~\ref{fig:fornax_pops}). The other component is warm -- it
contains a significant number of A/F dwarf or sub-giant stars. This
component could have several origins but only two are considered
here. Each has different consequences for K-band spectral index
behavior.

First, the warm component could be associated with a young population
with a warm ($\sim$2 \msun) main sequence turnoff (MSTO). The MSTO
stars are tied to thermally pulsating asymptotic giant branch (TP-AGB)
stars with bolometric magnitudes that place them above the tip of the
first-ascent red giant branch (TRGB). The TP-AGB stars are cooler than
stars on the first-ascent RGB and hence have an M spectral type. In
the underlying luminosity function (spectral type vs. number of
stars), the M-star bins are relatively more populated than in a purely
old galaxy. The net effect is that integrated 2.2 $\mu$m spectra
should become more M-like.

Second, the warm component could be associated with a metal-poor
population with a warm MSTO. These MSTO stars are tied to RGB stars
with similar luminosity as the corresponding metal-rich RGB stars but
with warmer \teff. In the underlying luminosity function, the K-star
bins will be relatively more populated than in a purely old
population. The net effect is to make the integrated 2.2 $\mu$m
spectra more K-like.

%
%
\begin{figure}
\resizebox{\hsize}{!}{\includegraphics[angle=0]{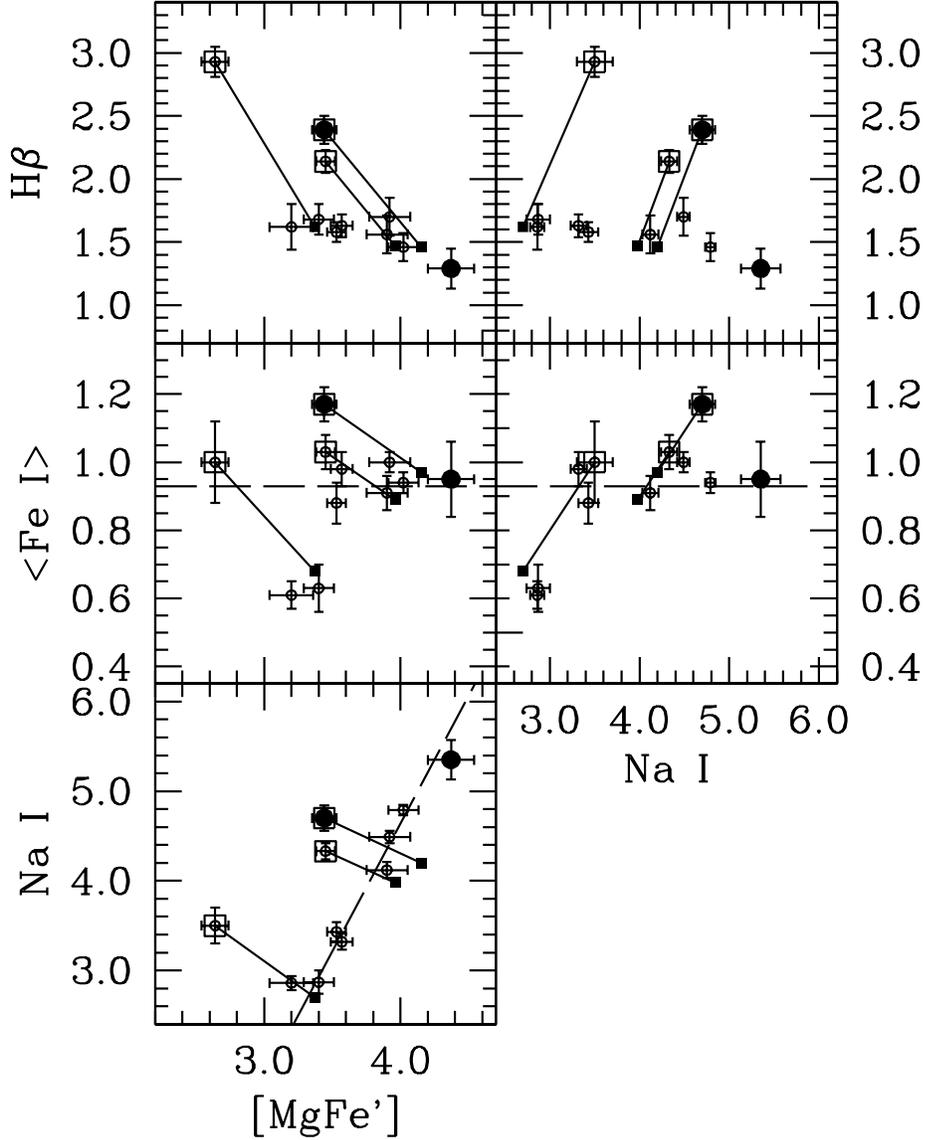}}
\caption{\label{fig:galModel} Optical-NIR corrections for young
populations.  See Figure~\ref{fig:galLineSigma} for a symbol
explanation. The observed index strengths for the galaxies with young
components are connected to the predicted index strengths ({\it solid
squares}) after the effects of a young component has been removed. In
the lower panel, the {\it slanted dashed line} indicates the
unweighted linear regression fit for the purely old galaxies with
$\sigma >$ 150 \kms. In the middle panels, the {\it horizontal dashed
line} indicates the unweighted mean value for purely old galaxies with
$\sigma >$ 150 \kms.}
\end{figure}

Can the observations presented here distinguish between these two
scenarios?  Consider Figure~\ref{fig:galModel} -- the galaxies with
relatively strong \hb\/ and weak \mgfe\/ have relatively stronger
(i.e. more M-like) near-IR features.  Qualitatively, this is
consistent with the first scenario. Can this conclusion be better
quantified? What is the relative ratio of young to old stellar mass?
Are the changes in optical and near-IR spectral features {\it
quantitatively} consistent?

To begin to answer these questions, the Thomas et al. (2003) models
were used to create very simple two-component models that produced
optical index values that matched the observed values in the three
galaxies with strong \hb. One component was old (11 Gyr) with
metallicity set to [Z/H] $=$ 0.35 for NGC 1316 and NGC 1344 and 0.0
for NGC 1375. The other component had the same metallicity but a young
(1 Gyr) age. The relative mass fractions of these components were
varied until the model \hb\/ strength matched the observed \hb\/
strength. The resultant young mass fractions (where total mass $=$ 1)
were 0.135, 0.075, and 0.140 for NGC 1316, NGC 1344, and NGC 1375,
respectively. The implied differential correction between the purely
old model and the two-component model was then applied to the observed
data (see Figure~\ref{fig:galModel}).

Next, observed \fe\/ strength was adjusted manually until all three
galaxies lay within the locus of purely old galaxies with similar
$\sigma$ in the \fe\/ vs. \naK\/ panel. As part of this adjustment,
$\Delta$(\fe)/$\Delta$(\naK) was forced to agree with the value (0.62)
determined for the observed Galactic cluster giant stars (see
Figure~\ref{fig:starLineLine}). This is a trend in effective
temperature that is relatively insensitive to metallicity. Cooler
giant stars (more late K and early M like) have strong K-band spectral
features.

In addition to the $\Delta$(\fe)/$\Delta$(\naK) measured from the
Galactic open cluster stars, two other key numerical trends can be
computed:

\begin{eqnarray}
\Delta(\fe)/\Delta(\naK)   & = & 0.62 \\
\Delta(\naK)/\Delta(\mgfe)  & = & 0.70 (1.10) \\
\Delta(\fe)/\Delta(\mgfe) & = & 0.28 (0.44)
\end{eqnarray}

NGC 1316 and NGC 1344 could be forced to have consistent
trajectories. However, NGC 1375 (values shown in parenthesis) appears
to follow somewhat different trajectories. Obviously, this kind of
cartoon model is illustrative only and surely hides a plethora of
details. Indeed, the real trajectories are unlikely to be linear.
Nevertheless, a clear astrophysical conclusion emerges: for the
\hb-strong galaxies, the observed optical and near-IR features all
change in concert, consistent with a warm MSTO tied to an extended
RGB, which is in turn consistent with the presence of a young stellar
component, not a metal-poor stellar component.

\subsection{Galaxies with bright K-band SBF}

Liu, Graham, \& Charlot (2002) found that relative I-band and K-band
surface brightness fluctuation strength (expressed as an SBF color,
$\bar{I} - \bar{K}$) was essentially constant for the majority of the
early-type galaxies they studied in Fornax. However, two galaxies in
common with our sample (NGC 1419 and 1427) were found to have brighter
(larger) K-band SBF relative to the mean relationship (see also Mei et
al. 2001). A larger $\bar{K}$ is thought theoretically to arise from
the presence of an extended giant branch, i.e. more cool, bright, M
giants.

In the spectroscopic data discussed here, there are no indications of
such extended giant branches in these galaxies -- their central line
indices are consistent with an integrated stellar populations
dominated by an old, metal-rich population. Of course, the SBF
measurements were made in the outer regions of these galaxies, as
opposed to the optical and near-IR observations of the central regions
discussed here. It may be that the integrated luminosity of such a
component (if present at all) is not high enough in the central region
for detection by our method.

\subsection{Purely old galaxies}

In purely old galaxies, we have seen that:
\begin{itemize}

\item{} \naK\ is stronger than in Galactic open clusters and is highly
correlated with $\sigma$ and \mgfe.

\item{} \fe\/ saturates for $\sigma \gtrsim 150$ \kms.

\item{} \co\ is somewhat correlated with $\sigma$ and \mgfe, while
\caK\/ is not (unless NGC 1399 is not included in the regression fits,
see footnote 2 above).

\end{itemize}

The observed K-band spectral features are named for the dominant
elemental species in the {\it solar} spectrum. However, at the
effective temperatures of interest here, the on-band index definition
and their companion off-band continuum bands contain lines from other
elemental species as well (as Ramirez et al.~1997 discuss
comprehensively). By referring to the high resolution spectra of WK96,
the contribution from these other absorbers can be investigated and
used to explain the overall spectral feature behavior. In turn,
underlying astrophysical parameters related to galaxy formation and 
evolution are revealed.

\subsubsection{\naK\/ index}

In the current galaxy sample, \naK\/ is significantly stronger than
observed in the Galactic open cluster stars.  The off-bands defined
for the \naK\/ feature are relatively line-free, so we can focus on
the on-band spectral region (see the $\lambda$ Dra, M0 III spectrum in
WK96, p. 352). In addition to sodium, Sc, Si, and (to a lesser extend)
V are significant absorbers in this region. Origlia et al. (1997)
argued that [Si/Fe] is super-solar in a small sample of early-type
galaxies based in measurements of the Si I 1.59 $\mu$m feature.
Trager et al. (2000) have argued that both [Na/Fe] and [Si/Fe] are
super-solar in early-type galaxies based on models of optical spectral
feature behavior. The K-band \naK\/ measurements presented here are
consistent with enhanced silicon in the observed Fornax galaxies and
hence indirectly confirm the conclusions of Origlia et al. and Trager
et al.

\subsubsection{\fe\/ index}

The \fea\/ feature contains approximately equal absorption
contributions from Fe, Sc and Ti while \feb\/ is dominated by Fe
absorption features with some additional Sc absorption (in WK96, see
the $\lambda$ Dra, M0 III spectra on p. 346 and 348). Hence, the \fe\/
index (a combination of \fea\/ and \feb) is dominated by absorption
lines from Fe-peak elements). For purely old galaxies with $\sigma >$
150 \kms, \fe\/ is observed to remain constant while \naK\/ (an index
dominated by absoprtion lines from $\alpha$-elements) and \mgfe\/ (a
total metallicity [Z/H] indicator) become stronger as $\sigma$
increases. It is tempting to conclude that above $\sigma >$ 150 \kms\/
luminosity-weighted mean total metallicity [Z/H] continues to increase
as a function of central velocity dispersion, driven by a relative
increase of [$\alpha$/H], while [Fe/H] remains constant.

However, within the narrow [Fe/H] range of observed Galactic open
cluster stars, \fe\/ is more strongly correlated with effective
temperature than cluster [Fe/H] (see
Figure~\ref{fig:starLineColor}). Increasing mean total metallicity
[Z/H] in the galaxies should correspond to cooler mean RGB effective
temperature and hence increased \fe. Recall that the relationship
between \fe\ and mean RGB effective temperature has already been
exploited above to explain the observed behavior of \fe\/ (and \naK)
in galaxies with young stellar components. The dilemma is clear: how
can a temperature-sensitive index like \fe\/ remain constant while
mean RGB effective temperature decreases due to increasing [Z/H]?

Without appropriate population synthesis models or stellar spectra,
only a few speculative thoughts can be offered. In their detailed
comparison of non-solar aboundance population synthesis models with
measurements of spectral indices in the optical Lick system, Trager et
al. (2000) concluded that their best-fit models included enhancements
in C, N, Na, and Si (among others). As [$\alpha$/H] increases with
$\sigma$ (as suggested by increasing \naK), stronger CN bands in the
\fea\/ and \feb\/ on-band and off-band windows (see the WK96 spectra
referenced above and the band definitions in Table~\ref{tab:features})
may have the effect of depressing the local continuum and hence
decreasing the value of \fe\/ $=$ (\fea\/ $+$ \feb)/2. For example,
Trager et al. noted a similar effect from C$_{\rm 2}$ bands in the
optical Mg {\it b} index. Therefore, it may be that above some
critical mean [$\alpha$/H] and below some critical mean effective
temperature, \fe\/ stays within a narrow range and provides no useful
information about mean [Z/H], [Fe/H], or effective temperature.

A conclusive astrophysical explanation of \fe\/ in early-type galaxies
with $\sigma >$ 150 \kms\/ awaits larger galaxy samples, observations
of appropriate stars (e.g. in the Galactic bulge), and detailed
population synthesis models.


\subsubsection{\co\ index}

The \co\/ feature is dominated by the $^{\rm 12}$CO(2,0) bandhead and
has no significant absorption components from other elements. Like
\fe\/ above, \co\/ is highly correlated with $T_{eff}$ in the observed
Galactic open cluster stars and reaches larger values than 
observed in the galaxies. As a function of central velocity dispersion $\sigma$ 
in the observed galaxies,  \co\/ does not appear to saturate (or at least not as 
definitively as \fe). This adds credence to the suspicion that \fe\ is not tracing mean 
RGB effective temperature in galaxies with $\sigma >$ 150 \kms.  Based on 
the current measurements, no conclusions can be reached about relative 
carbon or oxygen abundance in these galaxies. 

\subsubsection{\caK\ index}

No obvious explanation presents itself for the behavior of \caK\/ in
this galaxy sample.  The \caK\/ feature is intrinsically complex,
consisting of contributions from many absorbers: Ca, S, Si, and Ti
(all $\alpha$-elements) as well as Sc and Fe (see $\lambda$ Dra, MO
III spectrum in WK96, p. 342). As a function of \teff\/ in the range
of interest, some of the contributing absorption lines get stronger
(Sc and Ti, the former faster than the latter), some features get
weaker (Si), and some remain roughly constant (Ca, S, and Fe)
(cf. Ramirez et al. 1997). Relative to solar abundance ratios, Trager
et al. (2000) have argued that Si and S are over-abundant and Ca is
under-abundant in early-type galaxies while [Ti/Fe] and [Sc/Fe] have
their solar values. In the Galactic open cluster stars observed here,
the \caK\/ feature is stronger at a given $J-K$ ($T_eff$) in the
solar-metallicity stars than in the more metal-poor stars.  Within the
galaxies observed here, \caK\/ appears to have significant scatter at
any given $\sigma$ or \mgfe.  No obvious explanation for this scatter
(or the global behavior of \caK\ with $\sigma$ and \mgfe) presents
itself at this time.

\section{Summary}
\label{sec:summary}

Using new, moderate resolution ($R \sim 2500$) K-band spectra,
spectral indices have been measured in the central regions of eleven
early-type galaxies in the nearby Fornax cluster. Based on these
measurements, the following conclusions were reached:

\begin{enumerate}

\item{} The \naK\/ feature is much stronger in these Fornax early-type
galaxies than observed in solar-metallicity Galactic open cluster
stars. This is attributed to relative [Si/Fe] (and possible [Na/Fe]) differences
between the open cluster stars and the Fornax galaxies, i.e. both are
larger in the Fornax galaxies than in the cluster stars.

\item{} In various near-IR diagnostic diagrams, galaxies with optical
indices indicative of a warm stellar component are clearly separated
from galaxies dominated by colder, presumably old ($\geq$ 8 Gyr)
stellar populations. Changes in the near-IR spectra features are
consistent with the presence of an cool component dominated by late K
and/or early M giants stars. In combination, the optical and near-IR
observations are consistent with the presence of a young stellar
component with a warm MSTO and a significant extended giant branch
consisting of TP-AGB stars.

\item{} For detecting a young stellar component, the \naK\/
vs.~$\sigma$ or \fe\/ vs.~$\sigma$ diagnostic diagram seems as
efficient as using H$\beta$ vs. \mgfe\/ (or other similar combinations
of optical indices).  The near-IR features have the additional
advantage that no emission-line correction is needed (as it is
necessary for the stronger optical Balmer lines).

\item{} The \fe\/ index saturates in galaxies with central velocity
dispersion $\sigma$ $>$ 150 \kms\/ dominated by old ($\geq$ 8 Gyr)
stellar populations. For $\sigma >$ 150 \kms, these Fe features are
unlikely to be useful for investigating stellar population differences
between early-type galaxies. Above $\sigma >$ 150 \kms, the continued
increase in \naK\ (and \mgfe) strength with $\sigma$ coupled with
constant \fe\ presents an astrophysical challenge. Although it is
tempting to conclude that [Fe/H] reaches a maximum value, while
[$\alpha$/H] (and hence [Z/H]) continues to increase, more
observational and population synthesis work is needed to understand
conclusively the behavior of \fe\/ in these high-mass early-type
galaxies.

\end{enumerate}

Adding these near-IR indices to the standard diagnostic toolkit for
analyzing the integrated light of early-type galaxies clearly has
great potential. To develop this potential, three obvious steps are
needed: observe more galaxies over a larger range of central velocity
dispersion (including field galaxies with existing optical data) and
extend current population synthesis models to the near-IR. We know
that various groups are working on both steps.  It would also be
useful to study the radial behavior of these indices within individual
galaxies to compare and contrast index behavior between galaxies.

For the foreseeable future, the study of the central populations in
early-type galaxies will remain the study of integrated light. As
distance increases, our ability to study the central regions of
early-type galaxies within metric apertures equivalent to nearby
galaxies relies on the high {\it spatial} resolution spectroscopy
achievable from space or with adaptive optics on the ground. In the
former case, the James Webb Space Telescope represents the frontier --
yet, no optical spectrograph that works below 0.8 $\mu$m is currently
planned. In the latter case, the frontier will be shaped by further
development of adaptive optics systems -- and these systems will
achieve their best performance beyond 1 $\mu$m. Hence, understanding
how to interpret near-IR spectral indices in nearby galaxies is key to
facilitating the kind of investigation and characterization of more
distant early-type galaxies that will not be possible at optical
wavelengths.

\acknowledgements

We thank the anonymous referee for a quick, enthusiastic, and helpful
review that drove us to clarify several key points. 
This research has made use of the NASA/IPAC Extragalactic
Database (NED) which is operated by the Jet Propulsion Laboratory,
California Institute of Technology, under contract with the National
Aeronautics and Space Administration. This research has also made use
of the SIMBAD database, operated at CDS, Strasbourg, France. Thanks to
Valentin Ivanov and Olivier Marco for observing assistance during the
acquisition of the ISAAC data.  DRS warmly thanks ESO Santiago for
hospitality during July 2003 science leave, especially Danielle
Alloin, during which much of these data were processed.  We
acknowledge various helpful discussions with Andrew Stephans, Daniel
Thomas, Claudia Maraston, Scott Trager and Guy Worthey on topics
related to stellar populations in early-type galaxies.  Thanks to
Chris Lidman and Rachael Johnson for various discussions about
reducing ISAAC spectroscopic data.  Many of the science and technology
concepts in this paper grew out of late night conversations between
DRS and Richard Elston (1961 -- 2004) -- friend, colleague, fellow
observer --- while we were both postdocs at Kitt Peak. Richard ---
thanks.

{\it Facilities:} 
\facility{VLT:Antu (ISAAC)}
\facility{VLT:Yepun (SINFONI)}

\end{document}